\documentclass[12pt,preprint]{aastex}

\slugcomment{accepted by ApJ, 2004 June 7}

\begin{document}


\title{Gamma-Ray Burst Polarization: Limits from RHESSI Measurements}


\author{C. Wigger,
        W. Hajdas, 
        K. Arzner,
        M. G\"udel,
       A. Zehnder}
\affil{Labor f\"ur Astrophysik, Paul Scherrer Institut,
    CH-5232 Villigen PSI,
    Switzerland}
\email{claudia.wigger@psi.ch, 
       wojtek.hajdas@psi.ch,
       arzner@astro.phys.ethz.ch,
       guedel@astro.phys.ethz.ch,
       alex.zehnder@psi.ch}



\begin{abstract}
Using the RHESSI satellite as a Compton polarimeter,  a recent
study claimed that the prompt emission of GRB021206 was
almost fully linearly polarized. 
This was challenged by a subsequent reanalysis.
We present an novel approach, applying our method to the same
data.
We identify Compton scattering candidates by
carefully filtering events in energy, time, and scattering geometry.
Our polarization search is based on time dependent scattering rates in 
perpendicular directions, thus optimally excluding systematic errors. 
We perform simulations to obtain
the instrument's polarimetric sensitivity, and 
these simulations include photon polarization.
For GRB021206,
we formally find a linear polarization
degree of $\Pi_{GRB}= (41 ^{+57}_{-44})$\%, concluding that the
data quality is insufficient to constrain the polarization degree
in this case.
We further applied our analysis to GRB030519B and found again
a null result.
\end{abstract}




\newpage
\section{INTRODUCTION}

\label{sec:intro}

One of the most outstanding problems in present-day models 
of the energy release in 
$\gamma$-ray bursts (GRB) relates to the conversion of the liberated 
energy into
observed electromagnetic radiation. In relativistic fireball models, 
the pressure of 
the photon-lepton plasma itself leads to relativistic expansion 
(e.g., \citealt{piran99}).
Baryons are then accelerated through their coupling to the electrons. 
At a more
basic level, the origin of the expanding fireball requires a definitive source
of energy release that may itself be a source for mass acceleration. 
\citet{woosley99}
and \citet{macfadyen99}  discuss the formation of relativistically expanding
jets from hyperaccreting, stellar-mass black holes that are 
formed as a consequence of
iron core collapse of a rotating, massive ($>30M_{\odot}$) star. 
Alternative models
for the extraction of rotational (disk and black hole) energy and 
its conversion
to expanding shells or jets involve magnetohydrodynamic processes, 
including
reconnection and dynamo operation in the accretion disks 
(e.g, \citealt{blandford77, 
galeev79, thompson96, katz94, katz97, meszaros97}).

The most direct evidence for the presence of magnetic 
fields in GRB comes from their
emission spectrum which is now widely interpreted in 
terms of synchrotron emission
from relativistic electrons. Whether there is a direct 
connection between
magnetic fields produced in the immediate environment 
of the GRB or the precursor
star and the fields that are, at much larger distances, 
responsible for the
observed synchrotron emission is not clear. 
\citet{medvedev99} argue that
the required field strengths {\it for GRB afterglow shocks} exceed those to be 
expected  from dragging a pre-existing progenitor field along the 
expanding shell, 
and that they also exceed field strengths that could be produced 
by compression of
interstellar magnetic fields. A field-generating mechanism
intrinsic to the shocks would thus be required. The situation is, 
however, less clear for
the prompt GRB emission.

The recent report \citep{CB2003} of strong linear polarization  
in the {\it prompt} 
$\gamma$-ray emission from GRB021206 \citep{grb021206} 
therefore stirred some excitement.
The observations were performed with the RHESSI satellite \citep{Lin02}.
From extensive modeling, the authors derived a polarization degree of  
$\Pi = (80\pm 20)$\%, 
compatible with the assumption of maximum polarization. 
The emitting electrons are commonly
thought to be accelerated in collisionless shocks by the 
Fermi mechanism to a energy
distribution, ${dN/dE} \propto E^{-p}$, with typically 
$p \approx 2$. 
In that case, standard synchrotron theory predicts that 
the intrinsic polarization degree 
$\Pi$ is a function of  $p$, namely $\Pi = (p+1)/(p + 7/3)$ 
\citep{rybicki79}. This 
value therefore constitutes a maximum for a homogeneous magnetic field.
For $p = 2$, one thus finds $\Pi_{\rm max} \approx 0.7$.
Tangled magnetic fields with different field vectors relative 
to the line of sight
will in general reduce $\Pi$.

The implications of these observations are far-reaching: 
Not only does this observation
further support the synchrotron radiation model, 
it also seems to require nearly
homogeneous magnetic fields over the source visible to the observer
(a solid angle of $\approx \Gamma^{-2}$).  
\citet{CB2003}  argue
that magnetic fields dragged by the expanding shell from the surface 
of the exploding 
object are too weak to produce the prompt GRB, requiring additional 
turbulent shock-generated
fields that, however, would produce unpolarized emission. Therefore, 
they suggest
that the magnetic fields produced in the central engine are 
responsible for driving
the fireball themselves. Alternatively, post-shock dynamos 
could generate the highly
ordered magnetic fields; but then it will have to be shown 
that the instabilities occur
on large spatial scales \citep{CB2003}. 
\citet{lyut03} calculated the pulse-averaged polarization degree
for a relativistically expanding shell that contains a global toroidal field.
The synchrotron emission is again assumed to be from power-law distributed
electrons. Their calculations predict a maximum of 60\% for the polarization
fraction, depending on the spectral parameters of the GRB.
This view is not unequivocal. \citet{waxman03} points out that a slight
off-axis orientation of the relativistically expanding jet produces a
strong linear polarization signal even in the presence of random magnetic
fields. For further views on this, we refer to the summary in Lyutikov et
al. (2003) and references cited therein.

In any case, the prospect of diagnosing magnetic field structures 
in the very emission region of the prompt GRB deserves upmost attention. 
\citet{RF2003} have reanalyzed the RHESSI data of GRB021206 
and cast serious doubts on the polarization measurements 
reported earlier by \citet{CB2003}.
\citet{RF2003} conclude that
the signal reported by \citet{CB2003} is either spurious or is not related to
polarization, and they claim that no statistically meaningful statement can
be made for the degree of linear polarization for this particular GRB.

We have reanalyzed GRB021206
and find problematic issues in both previous analyses. 
We find, in line
with \citet{RF2003}, clear evidence that much of the signal claimed by 
\citet{CB2003} is induced not by source polarization but by accidental
coincidences of two unrelated photons arriving at the same time
in two different detectors. 
We essentially
confirm the null results of \citet{RF2003}, 
but we present a polarization analysis that compares simultaneous
scattering rates of orthogonal detector pairs and thus does not need
a complicated normalization. 
Additionally, after implementing photon polarization in the
GEANT3 software, we made
extensive simulations of RHESSI's response to a fully polarized GRB.
We also add one further GRB to this analysis, again
finding no statistically significant evidence for non-zero polarization.

The structure of the paper is as follows: We start
with a description of the RHESSI satellite and its relevant
features for polarization analysis in \S \ref{sec:rhessi}.
Next, we present the RHESSI data of GRB021206 in  \S \ref{sec:data}.
In our data analysis (\S \ref{sec:method}), 
we first select coincidence events 
(\S \ref{sec:c_types}),
before we search for a polarization signal in \S \ref{sec:polanal}.
Then, in \S \ref{sec:simul}, we present our simulations of a fully polarized GRB.
The results of the polarization analysis of GRB021206 
are presented in \S \ref{sec:results}, together with
the results from our simulation of a fully polarized GRB.
After a quick look at GRB030519B - another candidate for polarization
analysis - we compare our analysis with previous
works (\S \ref{sec:comp}) and discuss in general the suitability of
RHESSI as a GRB-polarimeter in \S \ref{sec:req}.

\section{RHESSI INSTRUMENTATION}
\label{sec:rhessi}
The Reuven Ramaty High Energy Solar Spectroscopic Imager 
(RHESSI, \citealt{Lin02}) 
is a NASA Small Explorer (SMEX) mission that
was launched on 2002 February 5 into a
low Earth orbit (600 km altitude, $38\arcdeg$ inclination).
RHESSI's primary science goal is 
to study solar flares in X-rays and gamma rays by using high resolution
imaging spectroscopy.

The mission is equipped with a single instrument made up of
two main systems: an imager and a spectrometer.
  The imager \citep{Zeh02} consists of a 1.5$\,$m long tube 
  with a set of 9 grids mounted on the front end of the tube
  and an identical set mounted on the rear end.
  The system makes use of the rotational modulation principle 
to obtain images of solar flares \citep{Hur02}.
RHESSI rotates with a period of $T_{rot} \approx 4\,$s and
points always towards  the Sun.
  The spectrometer consists of nine large 
  germanium detectors cooled to 75$\,$K.
Only weak lateral shielding was applied around the Ge-detectors. 
This causes RHESSI to be an all sky monitor for photons 
with energies above roughly 50$\,$keV. 
Thus, in addition to solar observations, RHESSI is 
also suited for GRB observations.

\subsection{Germanium Spectrometer}

\begin{figure}
\plotone{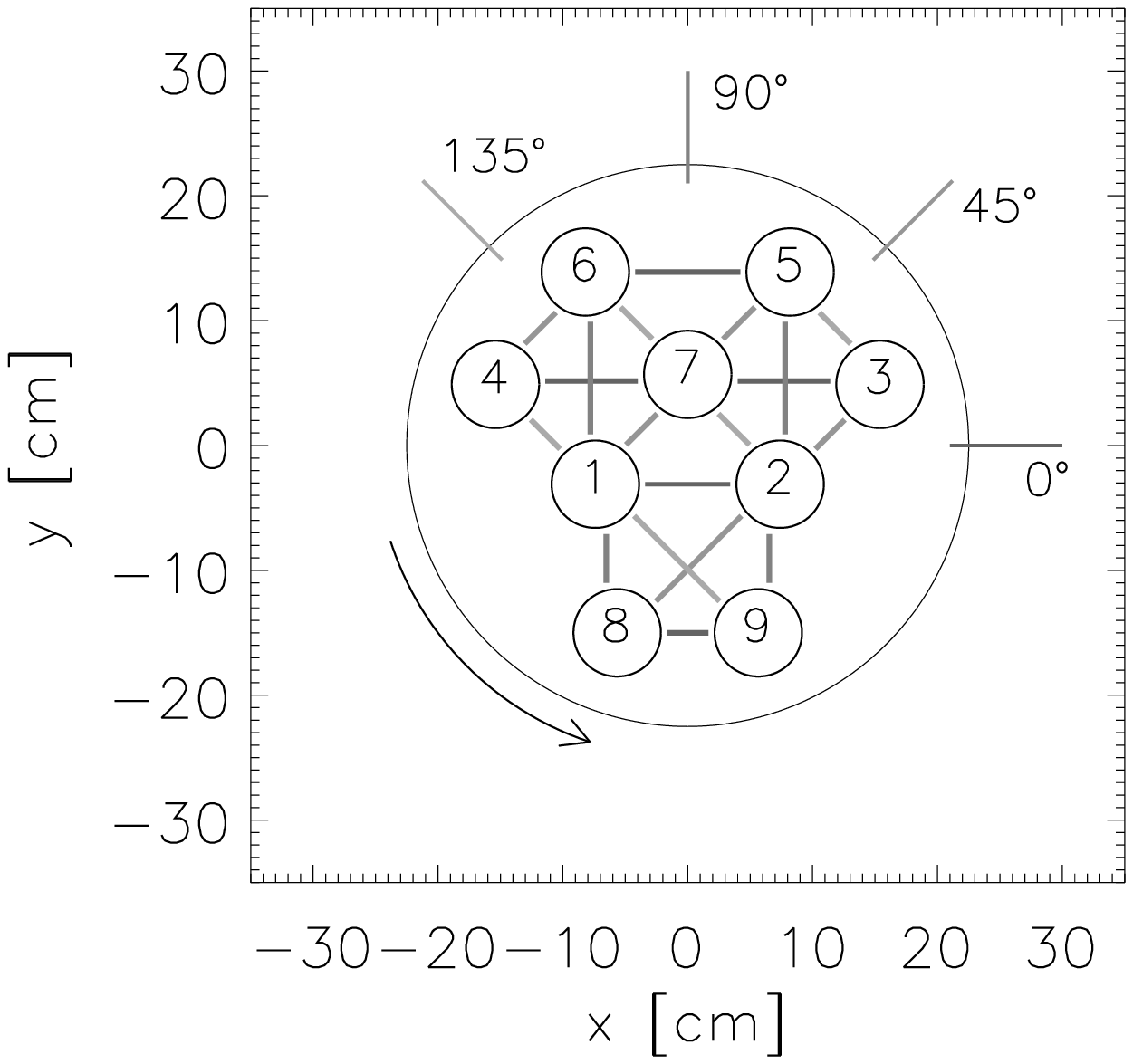}
\caption{Arrangement of RHESSI's 9 germanium detectors. 
  RHESSI spins around the axis perpendicular to the
  detector plane at 15$\,$rpm.
  The 19 ``close'' detector pairs a marked by bars.
  They can be grouped into four possible scattering directions.
  \label{fig:dets}}
\end{figure}

The arrangement of RHESSI's nine coaxial  
germanium detectors is shown in Fig.\ \ref{fig:dets}.
They have 
a diameter of 7.1$\,$cm and  a height of 8.5$\,$cm each.
The cylinders are electrically segmented in a thin front segment
(about 1.5$\,$cm thick) and a thick rear segment (about 7$\,$cm thick),
for details see \citet{Smith02}.
In reality the values of the front versus rear
thickness may vary significantly from detector to detector.
The front segments are able to detect photons from 3$\,$keV to 2.7$\,$MeV
with a resolution (FWHM) of $\approx$ 1$\,$keV (at 100$\,$keV), 
while the rear segments are sensitive from about 20$\,$keV to 17$\,$MeV
with a resolution of $\approx$ 3$\,$keV (at 1$\,$MeV).
A shielding plate
with holes for the detectors is placed on top of the cryostat 
which protects the
rear segments from solar photons scattered in the cryostat materials. 
Similar graded Z cylinders surround the front segments and protect them 
from low energy background photons ($E < 100\,$keV) 
coming from the side. 

\begin{figure}
\plottwo{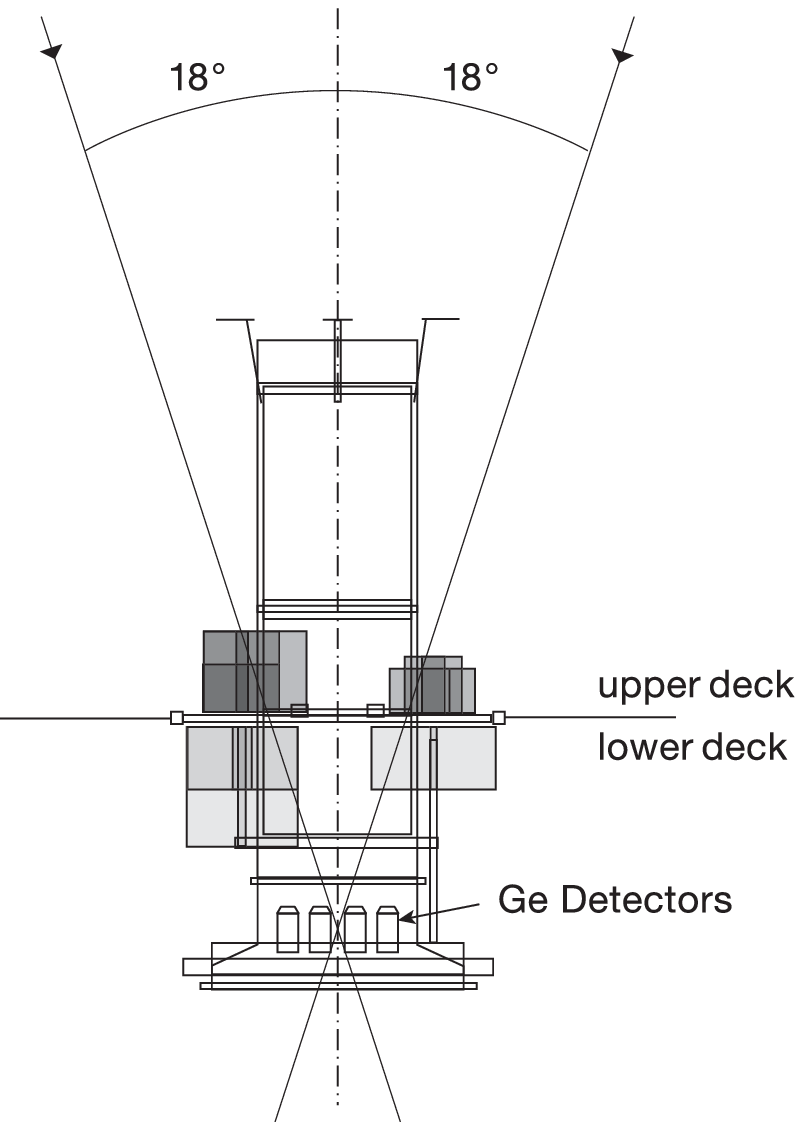}{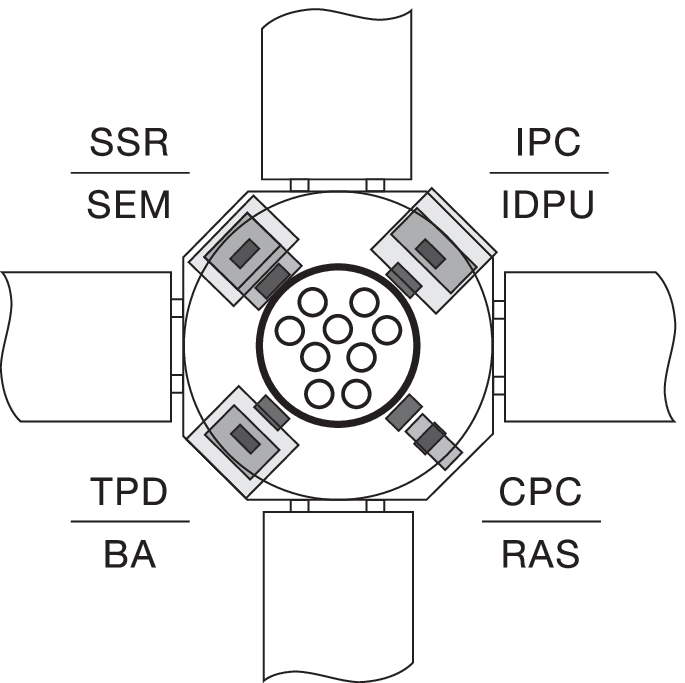}
\caption{Schematic view of the locations of the support modules. 
  On the solar direction of the equipment deck (i.e.\ above) are mounted:
  Solid State Recorder (SSR), 
  Transponder (TPD), 
  Cryocooler Power Converter (CPC), and  
  Instrument Power Converter (IPC).  
  The support modules mounted below the equipment deck are:
  Spacecraft Electronics Module (SEM), 
  Battery (BA), 
  photomultiplier Roll Angle System (RAS), and 
  Instrument Data Processing Unit (IDPU).
  Photons from GRB021206 came in at an angle of $18\arcdeg$ with
  respect to the Sun's direction. 
   \label{fig:hessi}}
\end{figure}

With respect to GRB observations, RHESSI's detection threshold
depends on the effective shielding
encountered by the incoming gamma rays. 
In the particular case of GRB021206, which 
was located at an angle
of $18\arcdeg$ with respect to the Sun's direction, the lower threshold
of the rear segments was about 100$\,$keV. 
This is mainly caused by 
absorption in the front segments and in the grid structures of the
imager, absorption in the shielding plate, and absorption in the
support modules located on the spacecraft equipment deck
(see Fig.\ \ref{fig:hessi}). 
These support modules are placed with a symmetry of $90\arcdeg$
in order to balance the space\-craft.
The uneven mass distribution of the support modules 
together with the rotation of RHESSI leads to
a periodic variability of the detection threshold, 
especially for GRBs coming in under such a small polar angle. 

\subsection{High Energy Polarimeter}
RHESSI was originally designed to measure linear polarization of hard
X-rays from the Sun with the help of a Be scatterer mounted in the plane
of the spectrometer \citep{McCon02}. Solar photons pass through 
predefined holes in the imager tube and,
after Compton scattering in the Be block, 
are detected in the neighboring rear segments. 
For Compton polarimetry of gamma rays coming from non-solar sources
(and therefore missing the holes for the Be scatterer), one rather
uses coincidence measurements between two detectors, one of which 
works as a scatterer. 

The azimuthal distribution of the Compton scattered
photons depends on the photon polarization: 
polarized photons are preferentially scattered in the direction
perpendicular to the initial polarization direction.
This azimuthal modulation is maximal for scattering angles
of about $90\arcdeg$.
Therefore, RHESSI's polarization sensitivity is highest for photons
approaching the satellite from directions perpendicular to 
the detector plane, i.e.
close to its rotation axis 
(as was the case for GRB021206). 
The coincidence rate of two detectors is a function of the angle between the
photon polarization direction and the line through the detector
centers. As this angle changes with the rotation of the satellite, the
coincidence rate may exhibit a characteristic modulation pattern. 
Its amplitude is proportional to the polarization degree,
and its period is $T_{rot}/2$.
The maximum amplitude that a 100\% polarized GRB would produce
is an intrinsic function of the instrument and is
obtained from Monte Carlo simulations.

\subsection{Spectrometer Electronics }

The detector signals 
are amplified by charge-sensitive preamplifiers before being
sent to the Instrument Data Processing
Unit (IDPU) for further analog and digital processing:
amplification, digitalization, time stamping etc.. There are eighteen
independent readout channels, one for each detector segment. Each
detected event is defined by its detector number and segment,
its arrival time and its energy channel. 
The event time is measured in  
units of binary microseconds ($1\,$b$\mu$s$ = 2^{-20}\,$s).
The absolute time is known with an accuracy of a few milliseconds. 
The data are stored event by event in the onboard memory. 

At the time of the GRBs studied here,  
front and rear segments of detector 2 were coupled. 
Its energy resolution is
therefore much worse ($\approx 8\,$keV at $100\,$keV).
The readout of detector 2 was provided only via the front channel. 

In order to  
save onboard memory, an automatic ``decimation'' procedure is
applied to events in the rear segments: 
Only one out of $N_{d}$ rear events is stored
for energies below a predefined threshold.  
This rear decimation is switched on mainly 
when the spacecraft is at high magnetic latitudes, 
where electron precipitation and cosmic ray
background are high. 
This was the case during GRB021206 with
$N_{d}=4$ and energy channels below 1024 ($\approx 380\,$keV).

\section{RHESSI Data for GRB021206}
\label{sec:data}
We developed our own routines for the polarization search.
The basic data structure we use is 
the event list that contains for each
detected photon its arrival time, its energy and 
the corresponding detector segment number.
The basic analysis steps are described below.

\subsection{Light Curve and Raw Energy Spectrum}
 
\begin{figure}
\plotone{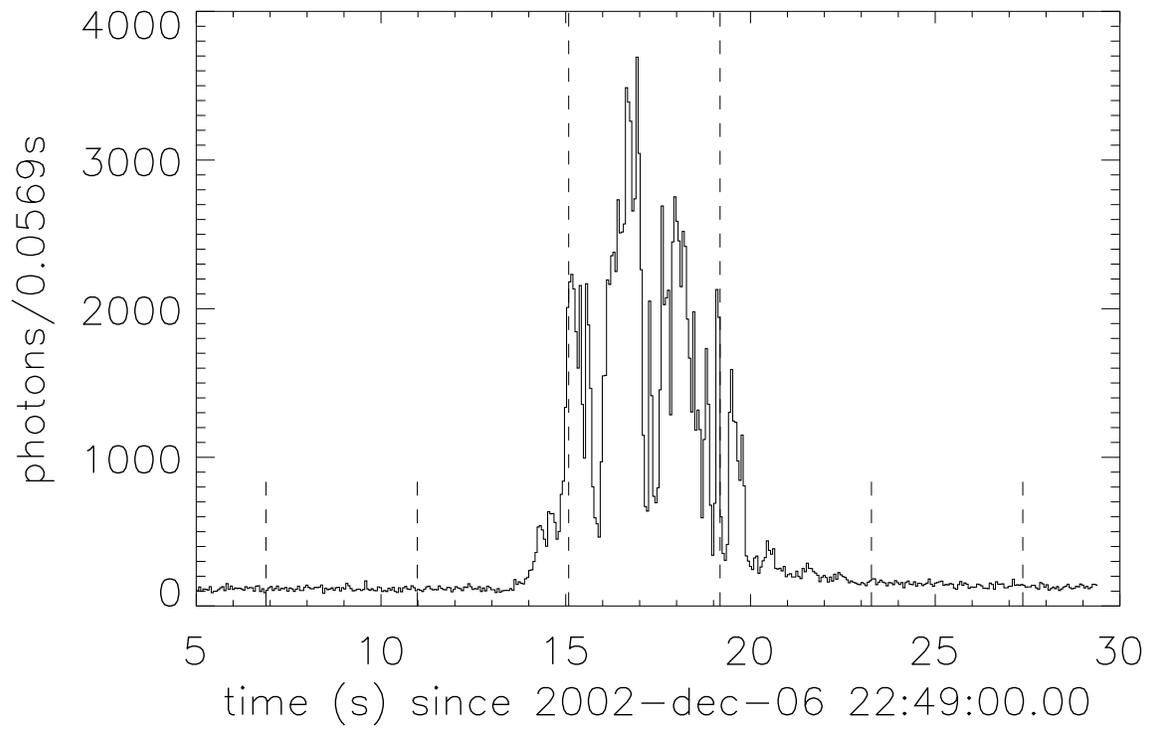}
\caption{Light curve of GRB021206 in the 25 -- 2000 keV band.
  Marked are the time intervals  
  used for polarization analysis (Eq.\ (\ref{t0_cut}))
  and for background determination.
  \label{fig:ltc}}
\end{figure}

\begin{figure}
\plotone{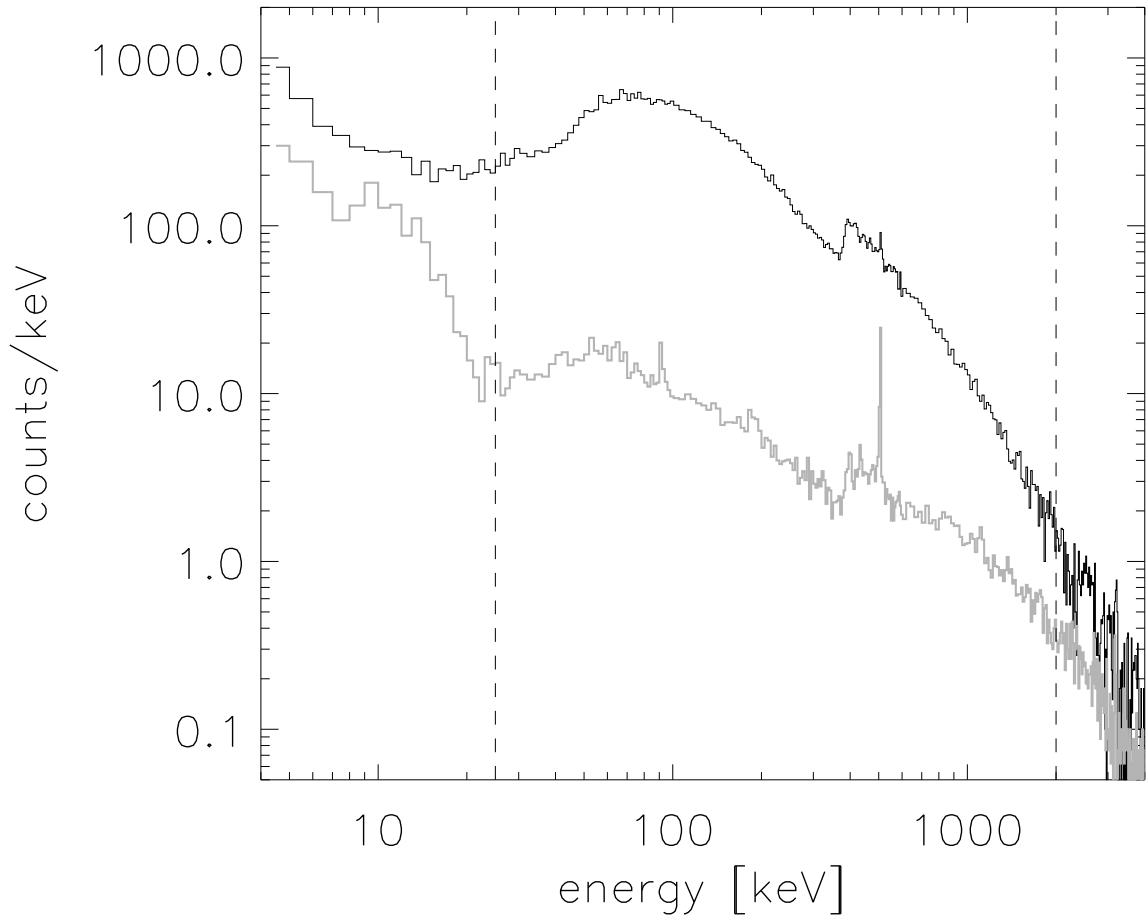}
\caption{Energy distribution in the time interval
  marked in Fig.\ \ref{fig:ltc}. Black: raw spectrum of all events.
  Gray: background spectrum, estimated from a time interval
  before and after the GRB.
  \label{fig:rspec}}
\end{figure}

The light curve of GRB021206, extracted from the event list
for all entries with energies
\begin{equation}
  25 \mbox{ keV} \leq \; E \; \leq 2000 \mbox{ keV} \label{eq:E0}  
\end{equation}
is shown in Fig.\ \ref{fig:ltc}.
For the polarization analysis, we selected a time interval that
corresponds to exactly one rotation of RHESSI 
(dashed lines in Fig.\ \ref{fig:ltc}):
\begin{equation}
 15.08 \mbox{ s} \leq \; t \; \leq 19.177 \mbox{ s ,}   \label{t0_cut}
\end{equation}
where $t$ is the photon arrival time since 2002 December 6 22:49:00 UT.
We find a total of about $130\,800$ photons in this time interval
and energy band.
The corresponding energy spectrum  is shown in
Fig.\ \ref{fig:rspec} (summed over all detectors and segments).
The step around 380 keV is due to decimation in the rear segments.
The 511 keV line originates from positron
creation and annihilation in the spacecraft and the atmosphere.

The total numbers of counts
in each detector segment are listed in Table \ref{tab:counts},
using the time and energy interval of Eq.\ (\ref{eq:E0}) and 
Eq.\ (\ref{t0_cut}). Detector 2 has no rear counts and 
a high total count rate since it is 
treated as an undecimated front segment.
The number of counts in the other detectors varies significantly, 
especially in the front segments, 
whose thicknesses are different from detector to detector.

\subsection{Definition of Coincidences}

\label{sec:coinc}
A photon that is Compton scattered from one 
detector into another one 
makes signals in two
detectors, and thus makes two coincident entries (indices $i$ and $j$)
in the event list.
Naively, one would expect the real coincidences to happen
within the same binary microsecond. It turns out, however, 
that real coincidences in the RHESSI data 
can have a time difference $dt=t_j-t_i$ up to $3\,$b$\mu$s (see below).

Compton scattering is the most important interaction of photons
with Germanium at energies from 150$\,$keV up to a few MeV.
The energy $k_{1}$ of a Compton scattered photon is
\begin{equation}
   k_1 = \frac{k_0}{1+(k_0/m_e)\cdot (1 - \mbox{cos}\, \theta)}
   \label{eq:Compton}
\end{equation}
where $k_{0}$ is the initial photon energy, $m_e$ the electron mass
and $\theta$ the scattering angle between in- and out-going photon.
The kinetic energy $E_e$ of the recoil electron is
\begin{equation}
   E_e=k_0 - k_1
\end{equation}
For a 150$\,$ keV photon 
that makes a Compton scattering with $\theta = 90\arcdeg$
the energy of the recoil electron is $E_e \approx 35\,$keV.
In our observation,  
low energy noise dominates below about 15 keV (see Fig.\ \ref{fig:rspec}). 
We therefore require for the energies $E_i$ and $E_j$ of a 
coincidence pair (``energy cut''):
\begin{eqnarray}
   25\, \mbox{ keV} & \leq & E_k\;, \;\;\;\;\;\;\;\;k=i,j  \nonumber \\
  150\, \mbox{ keV} & \leq & E_i+E_j \; \leq \;2000\, \mbox{ keV}  
        \label{E0_cut} 
\end{eqnarray}
If we assume that the total energy of the scattered photon
is measured in one of the Ge detectors, and the total
energy $E_e$ of the recoil electron 
in the other one, we can calculate
the scattering angle $\theta$ from the above equation (\ref{eq:Compton}),
namely
\begin{equation}
   \mbox{cos}\, \theta = 1 - \frac{E_e}{k_1}\cdot \frac{m_e}{E_e+k_1} \;.
\end{equation}
By taking into account
the geometry of the detectors and the space in between
we allow $\theta$ within $90\arcdeg \pm 45\arcdeg$. This
rather generous angle range
corresponds to $0.3 \leq 1-\mbox{cos}\, \theta \leq 1.7\,$.
For each event pair (energies $E_i$ and $E_j$) we can 
therefore apply an additional energy cut (``kinematical cut''), namely
\begin{eqnarray}
   0.3 & \leq &  \frac{E_i}{E_j}\cdot \frac{m_e}{E_i+E_j} \leq 1.7 \nonumber \\ 
       &\mbox{or}&    \label{e_special} \\
   0.3 & \leq &  \frac{E_j}{E_i}\cdot \frac{m_e}{E_i+E_j} \leq 1.7 \;\;.  
      \nonumber 
\end{eqnarray}
Because we do not know which detector observed the scattered photon
and which one the recoil electron, we have to test both combinations.

Concerning the accepted detector pairs, 
we distinguish between spatially {\em close} and {\em distant} 
detector pairs, see Fig.\ \ref{fig:dets}.
Among the 36 possible pairs we have 19 close and 17 distant pairs.
The chance of a real Compton scattered event between distant 
detectors is considerably smaller than between close detectors. 
For pairs like e.g. detectors 2 and 6, it is essentially zero  
because detector 7 is located in between.
Since the mean free path of a Compton scattered photon is
about 1 to 2$\,$cm, it would be stopped or scattered in detector 7.
Indeed, when the analysis was performed with the distant detector pairs only, 
an insignificant number of Compton scattering candidates
was found (see Section \ref{sec:distant}).
The acceptance of only close detector pairs is
called ``close pairs cut''.

\begin{figure}
\plotone{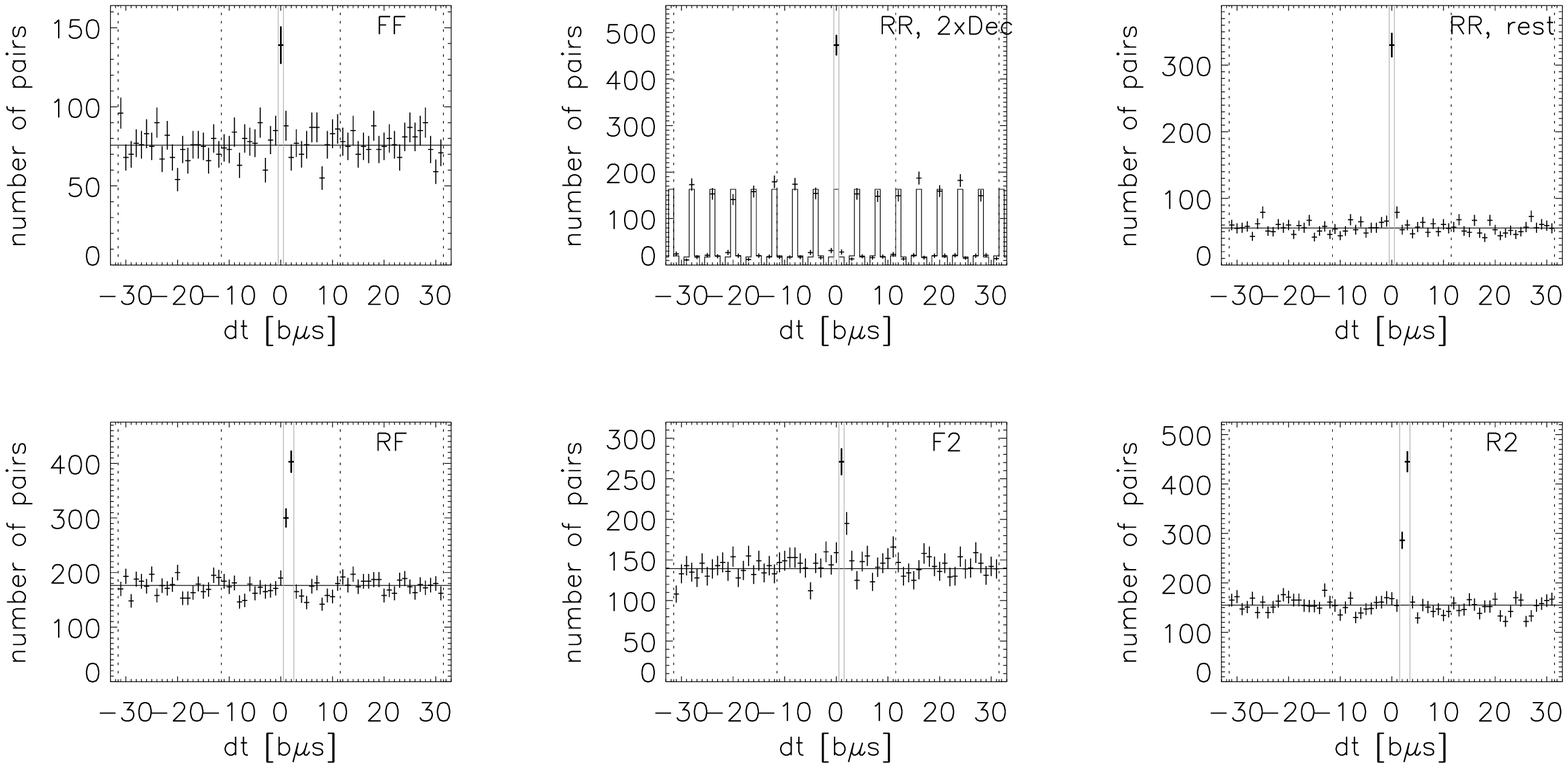}
\caption{Distribution of time differences $dt$ between any two photons
  in the event list (belonging to the time interval (\ref{t0_cut}),
  fulfilling the energy cut (\ref{E0_cut}), 
  the kinematical cut (\ref{e_special}), and the close pairs cut), 
  plotted for different combinations of electronics
  involved. The values in both intervals marked by 
  the vertical dotted lines were used
  to interpolate the distribution around $dt=0$. 
  The interpolation is illustrated by a solid line.
  Real coincidences appear only in a few bins with 
  0b$\mu$s $\leq dt \leq$ 3b$\mu$s. (Multiple coincidences are 
  included in these plots.)
  \label{fig:dtplot}}
\end{figure}

In order to find the time difference 
$dt$ of {\em real} coincidences,
we study the number of pairs as a function of $dt$ 
for {\em any} two events 
(not only two consecutive entries in the event list)
fulfilling (\ref{t0_cut}) and (\ref{E0_cut}), 
and occurring in a pair of close detectors. 
Fig.\ \ref{fig:dtplot} shows  six plots 
which represent the possible combinations between the involved 
electronics ({\tt F}: front segments, {\tt R}: rear segments, and
{\tt 2}: the unsegmented detector 2).
{\tt RF}, for instance, stands for pairs for which one signal
occurred in a rear segment
and the other in a front segment, and
$dt$ is defined as the time difference between front and rear. Negative 
$dt$-values indicate cases where the front signal occurred earlier than
the rear signal. 
In order to obtain positive and negative time differences
in the cases of {\tt RR} or {\tt FF} pairs, we arbitrarily 
choose to subtract the time of the
detector with the lower number from the time
of the partner with the higher number. 
This does not affect our further analysis.
{\tt RR,2xDec} stands for coincidences between two rear segments, 
where both energies $E_i$ and $E_j$ are within the decimated range.

The structure in the $dt$-plot of the {\tt RR,2xDec}-pairs shows 
a period of $4\,$b$\mu$s, which is due to the decimation 
procedure with $N_d=4$.
Decimation in the rear segment works like a clocked veto \citep{cur99}: 
During $1\,$b$\mu$s,
the events in all detectors are accepted, during the next $(N_d-1)\,$b$\mu$s 
they are all rejected.

The real coincidences are clearly visible in Fig.\ \ref{fig:dtplot}.
We find that they appear 
at different time differences $dt$, mainly at:
\begin{eqnarray}
  &\mbox{\tt FF :}  & dt=0 \mbox{ b$\mu$s} \nonumber \\
  &\mbox{\tt RR :}  & dt=0 \mbox{ b$\mu$s} \nonumber \\
  &\mbox{\tt RF :}  & dt=1 \mbox{ b$\mu$s}\;\; \mbox{ or }
                  \;\;  dt=2 \mbox{ b$\mu$s}  \label{dt_cut} \\
  &\mbox{\tt F2 :}  & dt=1 \mbox{ b$\mu$s} \nonumber \\
  &\mbox{\tt R2 :}  & dt=2 \mbox{ b$\mu$s}\;\; \mbox{ or } 
                  \;\; dt=3 \mbox{ b$\mu$s .} \nonumber  
\end{eqnarray}
These time bins are indicated in Fig.\ \ref{fig:dtplot}
by the vertical gray lines, and they are used as ``$dt$-cut''.
Coincidences outside these time bins are accidental coincidences. 
We think that the different $dt$ are caused by different
delays in the readout channels:
The rear segment electronics
is fastest, the time stamp in the front segments is on average 
$1.6\,$b$\mu$s later, 
and detector 2 is about $2.6\,$b$\mu$s later.

\section{ANALYSIS METHOD}
\label{sec:method}

\subsection{Coincidence Types}
\label{sec:c_types}

The high photon rate during GRB021206 introduces numerous accidental
coincidences. 
Their rate can be determined from Fig.\ \ref{fig:dtplot}.
We use the accidental coincidences with 
$ 12\,$b$\mu$s $\leq |dt| \leq$ $ 31\,$b$\mu$s (so-called 
``time shifted coincidences'')
and interpolate them around $dt=0$ 
(see Fig.\ \ref{fig:dtplot}).
The time differences $|dt|$ used to
estimate the number of accidentals 
should on the one hand be higher than the 
longest readout time of the events, i.e., longer than a few
b$\mu$s. On the other hand, it should be smaller than
the shortest time scale of variability in the GRB light curve.

The total number of coincidences ($N_{tot}$) as well as the 
number of accidental coincidences ($N_{acc}$) 
can be determined for any time interval, 
leading to time dependent functions $N_{tot}(t_i)$ and 
$N_{acc}(t_i)$. The number $N_{acc}(t_i)$ is proportional
to the square of the total count rate, and therefore shows
a strong time variation.

Some of the real coincidences observed during the GRB
are induced by various background components. We call this number
$N_{BG}$. It could show a rotation angle dependent time variation 
and can be estimated by interpolation of 
the data before and after the GRB: 
\begin{eqnarray}     \label{eq:nBG}
N_{BG}(t_i) &=& 0.5 \cdot (N_{tot}(t_{i,1})-N_{acc}(t_{i,1}))            \\
           &+& 0.5 \cdot (N_{tot}(t_{i,2})-N_{acc}(t_{i,2})) \;\; , 
	    \nonumber
\end{eqnarray}			   
where $t_{i,1}$ denotes a 
time bin before and  $t_{i,2}$ a time bin after the GRB,
with equal rotation angle of RHESSI as at time $t_i$.

The remaining number of coincidences, namely
\begin{equation}   \label{eq:nC}
  N_{C}(t_i) = N_{tot}(t_i) - N_{acc}(t_i) - N_{BG}(t_i)  \;\;,
\end{equation}
are called ``Compton scattering candidates'' and 
includes the real Compton scattering events.

A closer look at all the close detector pairs that fulfill 
the energy cut (\ref{E0_cut}),
the kinematical cut (\ref{e_special}), and 
and the $dt$-cut (\ref{dt_cut}) shows
that many of them are part of a multiple coincidence
(3 or more detector segments firing at the same time).
It often happens that a photon undergoes a forward Compton scattering
in a front segment before being Compton scattered in the rear
segment into a neighboring detector. 
In other cases, 
several detectors observe a signal. 
This is relatively more often the case before and after the GRB.
Multiple coincidences can be caused by nuclear reactions, 
bremsstrahlung, or pair/photon cascades originating from 
the passage of high energy cosmic rays through the spacecraft. 
None of these multiple coincidences contribute to a polarization
signal. We therefore do not accept them.

We identify multiple coincidences by looking at all accepted
coincidences. Let $t_C$ be the time of a coincidence, more
exactly, the earlier time of the two events involved.
Two coincidences are members of a multiple coincidence
if $|t_{C,1}-t_{C,2}| \leq 2\,$b$\mu$s. All coincidences that fulfill
this condition are rejected
(so-called ``no-multiples cut'').
However, members of multiple coincidences are also
members of time shifted coincidences used
to determine the number of accidental coincidences.
In order to obtain the correct number of accidental
coincidences, we clean the initial
event list from members of multiple coincidences.
We then search again for real and time shifted
coincidences.

\subsection{Polarization Analysis}
\label{sec:polanal}

In order to search for a polarization signal we first note that all
close detector pairs 
can naturally be grouped into
four scattering directions (see Fig.\ \ref{fig:dets}),
each separated by $45\arcdeg$. 
According to a RHESSI-fixed coordinate system, we
call these directions
the $0\arcdeg$ direction,  the $45\arcdeg$ direction,
the $90\arcdeg$ direction, and the $135\arcdeg$ direction.
The true pair angles, measured from
detector center to detector center, vary slightly. 
The mean deviation from the nominal value is $\approx 3 \arcdeg$
with a maximum of $8.4 \arcdeg$.
This is much less then the width of scattering angles accepted by a 
detector pair, which is of the order $30 \arcdeg$.  
  
Assuming that polarized photons at a time
$t_0$ were scattered preferentially in the 
$135\arcdeg$ direction, then,
one eighth of a satellite rotation later,
they would be scattered preferentially in the $90\arcdeg$-direction.
After another rotation by $45\arcdeg$, they would be scattered in the
$45\arcdeg$-direction, and so on.

We determine the number of Compton scattering candidates 
for the 4 scattering directions and call them
$N_{C,x}$, with $x \in$ \{ 0, 45, 90, 135 \}.
For a GRB with a strongly varying light curve, such as GRB021206, 
the $N_{C,x}(t_i)$ are strongly correlated with the 
light curve. 
Let $n_x(t_i)$ be the normalized coincidence light curves,
\begin{equation}
  n_x(t_i) = \frac{N_{C,x}(t_i)}{\sum_i N_{C,x}(t_i)} \,\,,    
      \label{eq:norm_cs_ltc}
\end{equation}
where the sums are over the entire time interval used for polarization
analysis (typically one or more full rotations, e.g. Eq.\ (\ref{t0_cut})).
We can then define the following asymmetries: 
\begin{eqnarray}
  A_{135-45}(t_i) & = &
      \frac{n_{135}(t_i)-n_{45}(t_i)}
           {n_{135}(t_i)+n_{45}(t_i)}  \nonumber \\
  A_{90-0}(t_i) & = &
      \frac{n_{90}(t_i)-n_{0}(t_i)}
           {n_{90}(t_i)+n_{0}(t_i)}\,\,,    \label{eq:asym}
\end{eqnarray}
For an {\em unpolarized} GRB, we make 
the plausible assumption that
the $N_{C,x}(t_i)$ are approximately proportional to each other, i.e.\  
we assume that there exists a time dependent function $f(t_i)$
and factors $B_x$ such that we can write
$N_{C,x}(t_i) \approx B_x \cdot f(t_i)$. 
The factors $B_x$ take into account the efficiencies of the
four scattering directions, while $f(t_i)$ is closely related
to the light curve of the GRB.
In that case, the asymmetries (\ref{eq:asym}) are
always zero (within statistical errors). 
In the case of a {\em polarized} GRB, 
an additional sinusoidal time dependence is introduced:
\begin{eqnarray}
 \lefteqn{N_{C,x}(t_i) \approx B_x \cdot f(t_i)} 
           \label{eq:ltc_mod}  \\
   && \cdot (1 + \mu_{p} \cos(\omega t_i - \phi - \phi_x))\,\,. \nonumber
\end{eqnarray}
The observation of a non-zero modulation amplitude $\mu_{p}$ would be
an indication of a GRB polarization. 
Simulations show
that $\mu_{p}$ is of order $0.2$ for a $100\,$\% polarized GRB.
The expected periodicity of the modulation is 
$\omega = 2 \pi/(T_{rot}/2)$. 
The absolute phase $\phi$ is unknown,
and the relative phase shifts are fixed at
$\phi_{135}=0$, $\phi_{90}=\pi/2$,
$\phi_{45}=\pi$, and $\phi_{90}=3 \pi/2$.
From Eqs. (\ref{eq:norm_cs_ltc})-(\ref{eq:ltc_mod}), we obtain
\begin{eqnarray}
   A_{135-45}(t)&\approx&
      \mu_{p} \cdot \mbox{cos}(\omega t - \phi)    \label{eq:fitfun} \\
   A_{90-0}(t)  &\approx&
      \mu_{p} \cdot \mbox{cos}\left(\omega t - \phi - \frac{\pi}{2}\right) \,.
           \nonumber 
\end{eqnarray}

The definition of the asymmetries $A_{135-45}(t)$ and $A_{90-0}(t)$
uses only data. By making a fit, the 
best values for the observable modulation amplitude $\mu_{p}$
and the phase $\phi$ can be determined.
In order to obtain the real polarization degree of a GRB, $\mu_{p}$
should be compared with $\mu_{100}$ obtained from the
simulation of a 100\% polarized GRB, analyzed the same way.

\subsection{Simulations}
\label{sec:simul} 
The standard RHESSI software provides energy deconvolution routines,
using simulated response functions, 
but only for solar photons.
Therefore we have rerun the simulation code, but for off-axis photons.

All essential parts of the satellite are built 
into a mass model using the GEANT3.21 particle tracking 
code \citep{Geant3}.
The model provides an exact description of the
Ge spectrometer and its adjacent elements, e.g.\  graded-Z shielding,
detector electronics and cryo-cooler.
Other components and structures 
that are positioned at larger distances from the
Ge detectors, maintain their mass and composition while some
approximations were made for their geometry (e.g.\ simplified shapes) and
internal arrangement (e.g.\ using material of mean density and Z-number).
The total mass of the spacecraft is 291$\,$kg, including
131$\,$kg of scientific payload.

Especially for photons that arrive at an angle 
of less than  $30\arcdeg$  with respect to the Sun's direction 
(as was the case for GRB021206), 
the mass distribution on the equipment deck
affects the light curves and spectra.
The most relevant components responsible for such effects are
the solid state recorder 
   at $135\arcdeg$ above (i.e. Sun side of) the equipment deck, 
the cryo-cooler power converter 
   at $315\arcdeg$ (above the equipment deck), 
the IDPU at $45\arcdeg$ (below the equipment deck), 
and   
the battery cells at $225\arcdeg$ (below the equipment deck),
see Fig.\ \ref{fig:hessi} (or \citet{Lin02}, Fig.10).
They all weigh of the order 10$\,$kg. 

Using the mass model of the spacecraft and the spectrometer, the RHESSI
response function can be generated. For this
purpose the spacecraft model in the GEANT code is illuminated by a
parallel flux of gamma rays with a predefined energy spectrum and direction. 
All primary particles, like scattered photons, as well as secondary particles,
like recoil electrons, are traced.  
Finally, the energy
deposition in each germanium detector segment is stored.

Photon polarization is not included in GEANT3. We wrote specific
routines to fully include photon polarization,
using the same formalism as in the GEANT4 extension packet for low
energy Compton scattering \citep{Geant4, hei54}. 
The momentum vector of the
scattered photon and its polarization direction
were generated in accordance with the Klein-Nishina
formula taking into account its initial polarization.

In order to analyze the simulated data with the same program
as the real data, the output of the GEANT-code was prepared 
to look like real data. 
The effects of the readout electronics were simulated
by delaying the front segment events by 1.6$\,\mu$s and
the events in detector 2 by 2.6$\,\mu$s (see end of section \ref{sec:coinc}).
Next, the time was converted into integers (in units of b$\mu$s) and 
the energy was converted into channels.
The rear segment decimation was simulated by accepting  
rear segment events with energy channel $< 1024$ only if 
$(t \; \mbox{mod} \; 4) = 0$. 
We also added events as cosmic induced background.
In order to obtain good statistics, we simulated 
the GRB lasting over many rotations (typically 10 to 40).
A very high photon rate per rotation would introduce 
a very high accidental coincidence rate.

\section{RESULTS}
\label{sec:results}

\subsection{Results from GRB021206}


\label{sec:res_grb021206}
We start the data analysis of GRB021206 
by identifying possible coincidences.
Possible coincidences are
event pairs that have a time difference 
$dt \leq 3\,$b$\mu$s,
belong to the time interval given in 
Eq.\ (\ref{t0_cut}), and fulfill the energy cut (\ref{E0_cut}).
We find 16542 such pairs, see Table \ref{tbl:cut_eff}. 
Using the method described in  
section \ref{sec:c_types} to determine $N_{acc}$ and $N_{BG}$,
we find 1641 Compton scattering candidates.
The effects of successively applying the $dt$-cut, the
close pairs cut, the kinematical cut, and the no-multiples cut
are listed in  Table \ref{tbl:cut_eff}.
The $dt$-cut 
and the close pairs cut affect the number of 
Compton scattering candidates only slightly, but
are both very effective in reducing the number of accidental coincidences,
as can be seen from Table \ref{tbl:cut_eff}.
The kinematical cut seems less effective.
However, simulations show
that the modulation amplitude is about 15\% higher when 
using the kinematical cut. 
The sequence of the energy cut, the $dt$-cut, the close pairs cut
and the kinematical cut is interchangeable. 
The last line of Table  \ref{tbl:cut_eff} is obtained after
applying the no-multiples cut.
The no-multiples cuts is not interchangeable
with the other cuts.
The quite big effect of the no-multiples cut can be explained by the fact 
that multiple coincidences can lead to comparably many double coincidences.
A triple coincidence of, e.g., detector 2, rear segment 3 and rear
segment 5 can lead to three accepted coincidences. 
Multiple coincidences that include
4 or more detector segments will cause even more double coincidences.

\begin{figure}
\plotone{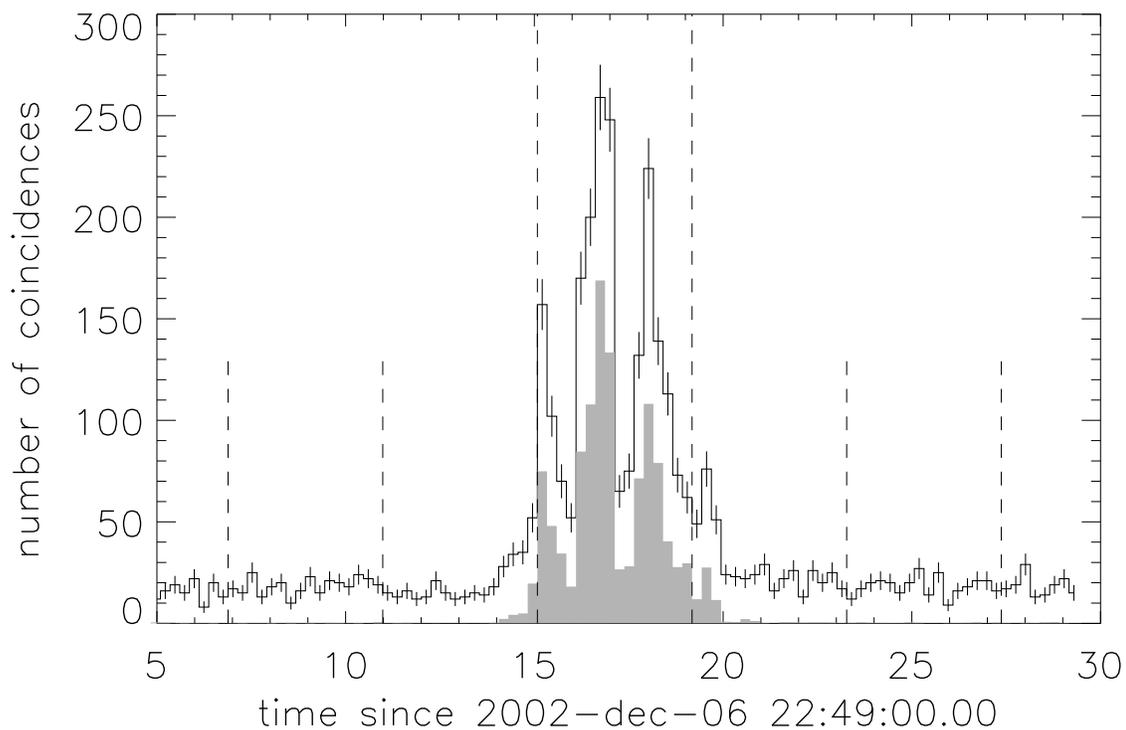}
\caption{Light curve of coincidences. 
  Black line histogram: total number of coincidences per bin ($N_{tot}$).
  Gray filled histogram: accidental coincidences ($N_{acc}$).
  The time bin width is $T_{rot}/16$.
  The long vertical lines indicate the time interval chosen
  for the polarization analysis. The short vertical lines indicate
  the intervals used to estimate the background contribution ($N_{BG}$).
  \label{fig:t1hist}}
\end{figure}

The time dependent total coincidence rate $N_{tot}(t_i)$ of GRB021206 is 
plotted in Fig.\ \ref{fig:t1hist} (crosses),
together with the number $N_{acc}(t_i)$ of accidentals 
(gray filled histogram). Both  $N_{tot}$ and 
$N_{acc}$ were determined as described in 
section \ref{sec:c_types}.
It can be seen that 
a large fraction of all coincidences during the GRB
are accidentals. Before and after the GRB, we observe
background coincidences.  
The number of accidental coincidences
is very small before and after the GRB.
      
      We note at this point that the energies 
      of the background
      coincidences are on average higher than
      those of the GRB coincidences. We also observed that
      the percentage of multiple coincidences is higher
      among the background coincidences than among the
      GRB induced coincidences.

We obtain a total of 
$N_{tot} =2141\, (\pm 46)$
coincidences in the time interval between
the two dashed long vertical lines of Fig.\ \ref{fig:t1hist},
and 
$N_{acc} = 1081\, (\pm 10)$ 
accidental coincidences  
(the error is $\sqrt{1081/10}$ 
because the interpolation of the accidentals 
relies on 10 times better statistics). 
From the time interval before and after the GRB (indicated by the short
dashed lines in Fig.\ \ref{fig:t1hist}) we determine the number of
background coincidences as 
$N_{BG} = 290 \, (\pm 12)$.
Thus we find 
\begin{equation}  \label{eq:N-C-grb021206}
  \mbox{GRB021206: \hspace{2em}}  N_{C} = 770 \pm 49 
\end{equation}
Compton scattering candidates.
The cited errors are purely statistical.

\begin{figure}
\plotone{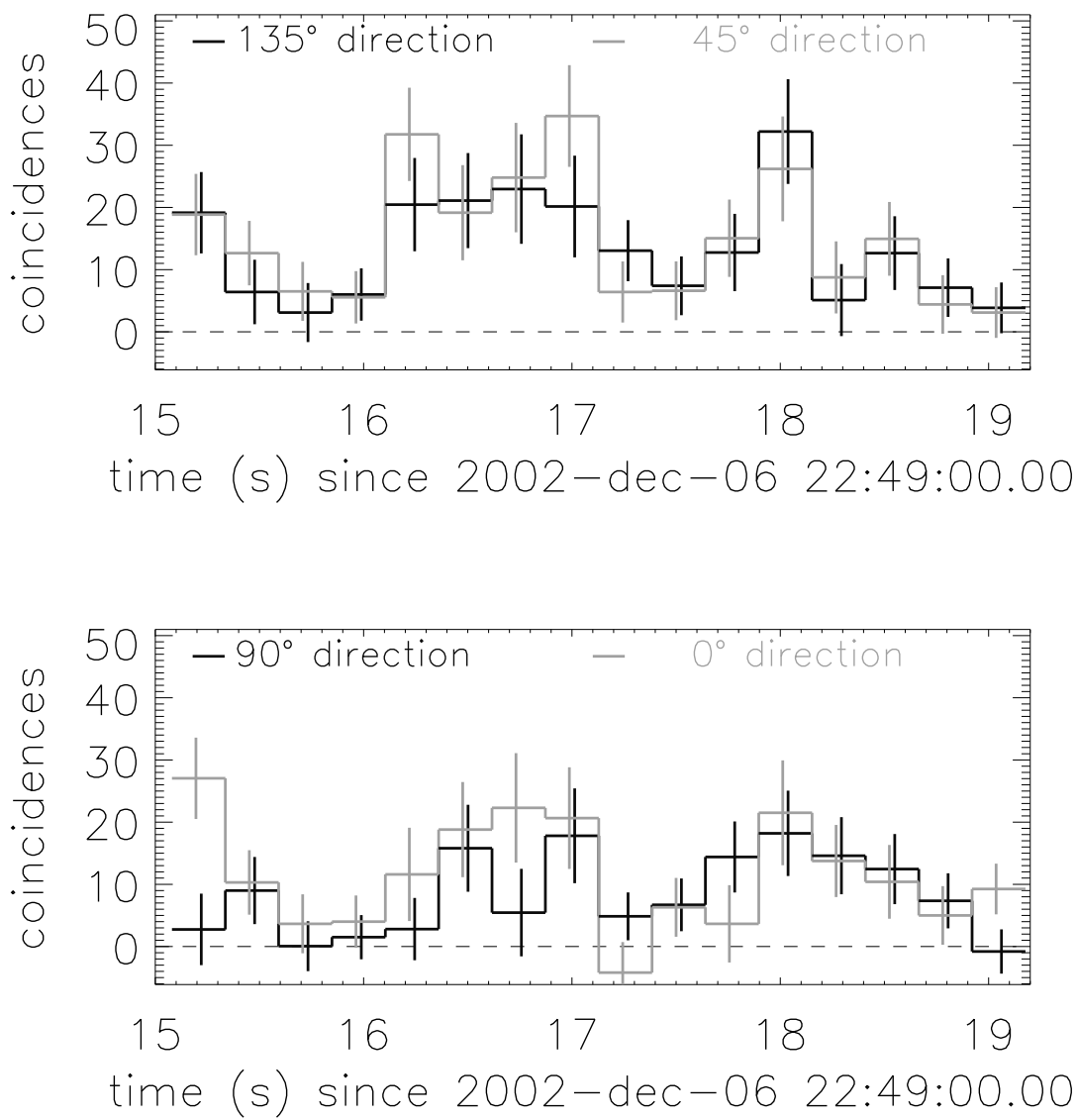}
\caption{Time distribution of Compton scattering candidates, for
    the 4 different scattering directions given by the layout 
    of the RHESSI detectors (see Fig.\ \ref{fig:dets}).
    The time bin width is 1/16th of a full rotation, 
    as in Fig.\ \ref{fig:t1hist}.
    Two orthogonal directions are 
    shown in the same plot. 
    If one curve of a plot -- in case of a polarized burst -- 
    showed a relative maximum, the other curve would have a relative
    maximum 4 bins later and earlier.    
   \label{fig:diffhist}}
\end{figure}

The time distributions of the Compton scattering candidates
during the GRB ($N_{C,x}(t_i)$) are plotted in
Fig.\ \ref{fig:diffhist} for the four possible scattering directions
(see section \ref{sec:polanal}). 
The two directions in each plot make an angle of $90\arcdeg$.
If, in case of a polarized burst,  
one curve showed a relative maximum 
compared to the other curve, 4 bins later ($\approx$ 1.0$\,$s)
the other curve would show a relative maximum.
Additionally, the two plots are related by the fact that a
relative maximum in the light curve of the $135\arcdeg$ direction 
(upper plot) would appear 2 bins 
($\approx$ 0.5$\,$s) later in the $90\arcdeg$ direction (lower plot).
No such effect can be seen by visual inspection.

\begin{figure}
\plotone{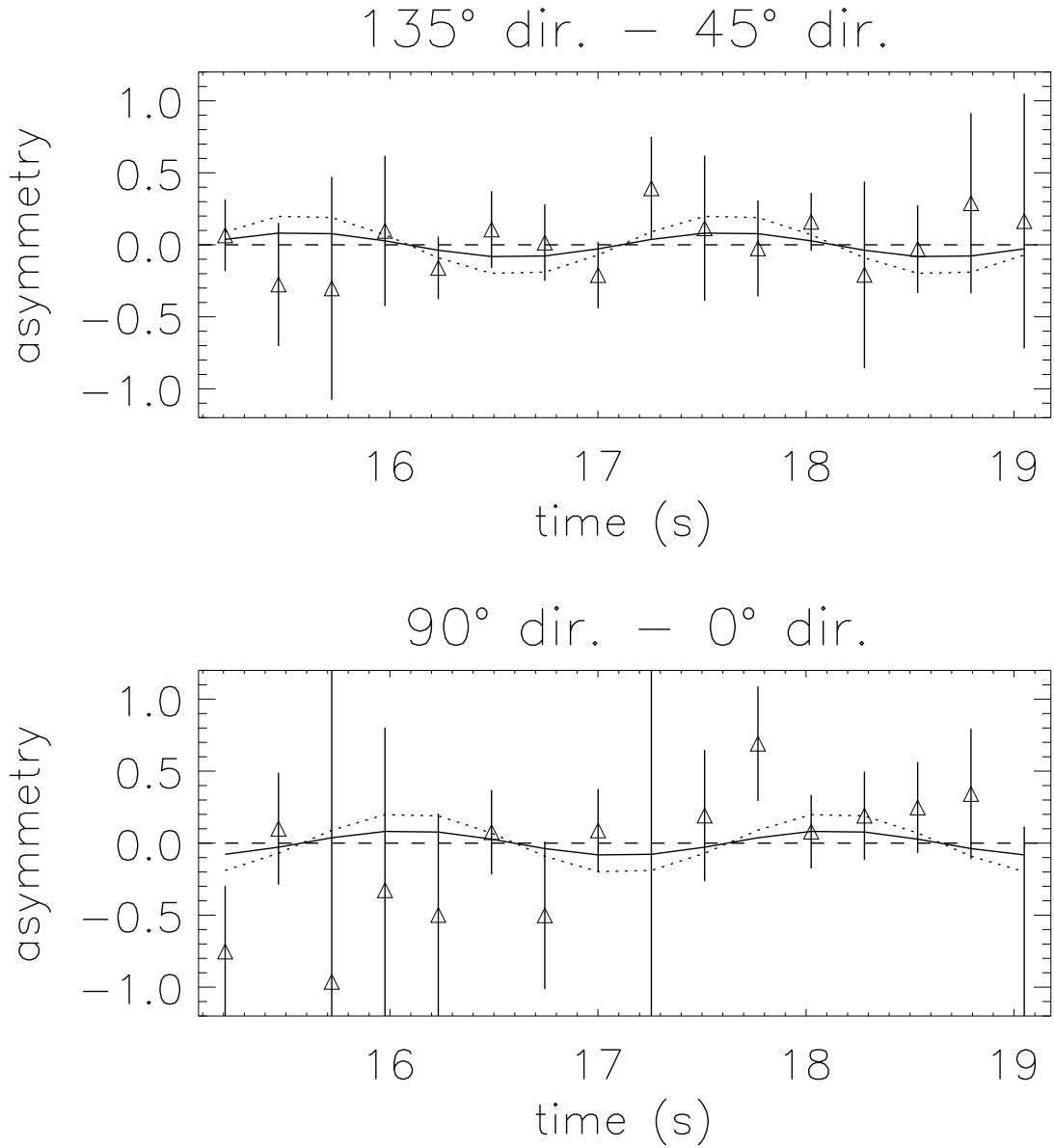}
\caption{Asymmetries of the curves shown in Fig.\ \ref{fig:diffhist},
   as defined in Eq.\ (\ref{eq:asym}).
   The two plots are fitted simultaneously.
   The black line is the best fit of a sine-function, see
   Eq.\ (\ref{eq:fitfun}), where the amplitude
   and phase are treated as free parameters.  
   The dotted line has an amplitude as
   a fully polarized GRB would make (see Eq.\ (\ref{eq:my100})). 
   Obviously, no significant
   statement about the polarization degree of GRB021206 can be made.
   \label{fig:asym}}
\end{figure}

The asymmetries $A_{135-45}$ and $A_{90-0}$, defined in 
Eq.\ (\ref{eq:asym}), are plotted in 
Fig.\ \ref{fig:asym}, together with the best fit of 
the function (\ref{eq:fitfun}).
Since the two curves are connected by a fix phase, 
we fit them simultaneously in order to obtain the two free parameters, 
namely the asymmetry amplitude $\mu_p$ and the phase $\phi$.
The fit result for the modulation amplitude is 
\begin{equation}   \label{eq:mp-grb021206}
  \mbox{GRB021206: \hspace{2em} } \mu_{p} = (8.6 \pm 9.4) \mbox{\%} 
\end{equation}
with a  $\chi^2$ of 14.0 (28 DoF).

\subsection{Results from GRB030519B}

We have also studied GRB030519B \citep{grb030519, IPN} 
as another good candidate
for polarization analysis.  
GRB030519B is also very strong and 
occurred at an angle of $165 \arcdeg$ 
with respect to the direction toward the Sun, 
i.e.\ it came from the antisolar side
at an anlge of only $15 \arcdeg$.
There was no decimation during this observation. 

With the method described in section \ref{sec:method}, but 
including two full rotations 
from  14:04:53.80 to 14:05:02.062 UT, 
we obtain a total of 
$N_{tot}=1471\,(\pm 38)$ coincidences, 
$N_{acc}=144\,( \pm 4)$ accidental coincidences, and 
the number of background coincidences 
is $N_{BG}=710\,( \pm 19)$, resulting in 
\begin{equation}  \label{eq:N-C-grb030519b}
  \mbox{GRB030529B: \hspace{2em}}  N_C = 617 \pm 43 
\end{equation}
Compton scattering candidates. 
More than half of the coincidences
originate from background.
The more than three times higher value for the background 
coincidences (compared with GRB021206)
can be explained by the two times longer GRB time interval, and 
by the undecimated data.

The fitted amplitude is 
\begin{equation}  \label{eq:mp-grb030519b}
  \mbox{GRB030529B: \hspace{2em}}  \mu_{p} = (4.7 \pm 8.6) \mbox{\%} 
\end{equation}
with a $\chi^2$ of
11.7 (28 DoF).

\subsection{Results from simulations}

\label{sec:simresult}
\begin{figure}
\plotone{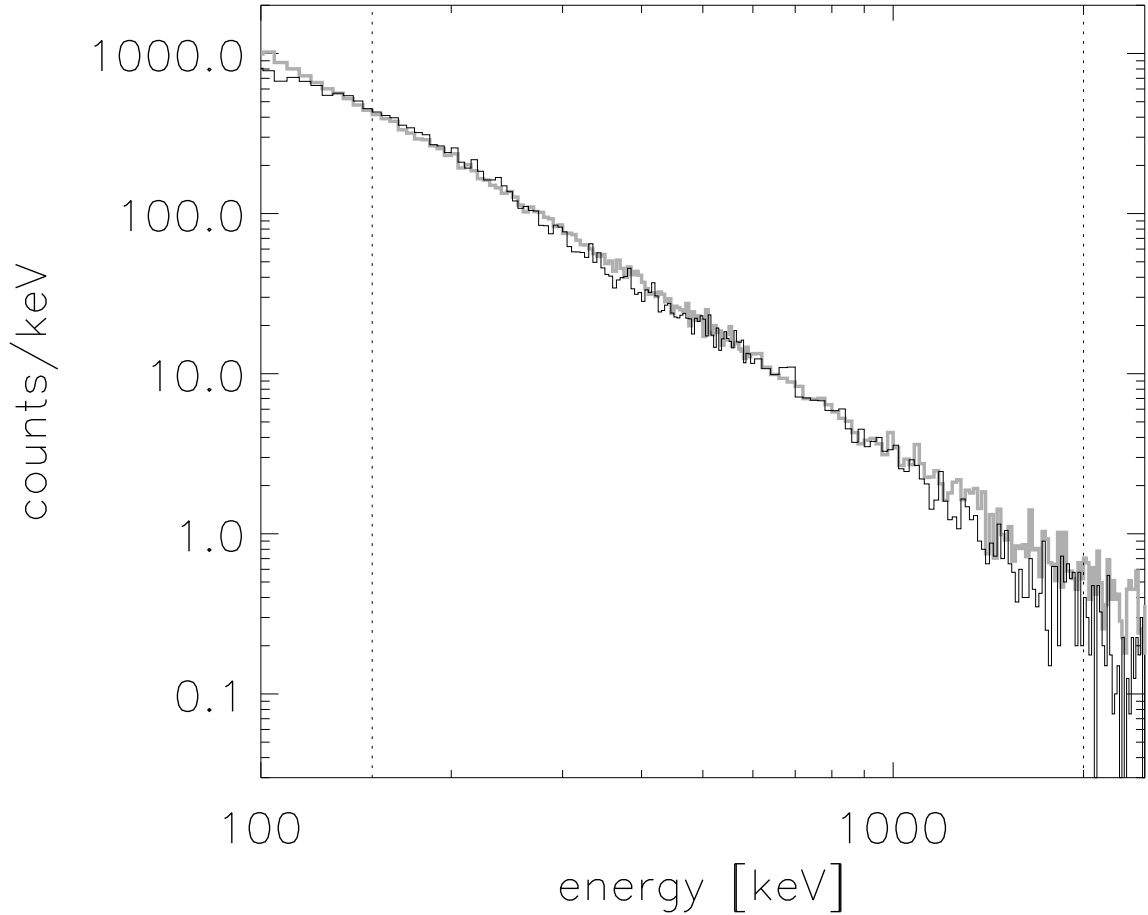}
\caption{Observed energy spectrum of GRB021206 (black), summed over
   all detectors and segments (except detector 2). 
   The background was subtracted, and the spectrum is corrected
   for decimation (unlike Fig.\ \ref{fig:rspec}).
   The gray histogram is the result of a simulation with  
   $dN/dE \propto E^{-2.6}\,$.
   \label{fig:comp_spec}}
\end{figure}

The initial energy spectrum is simulated between 25 keV and 5 MeV
with a spectral shape that follows a simple power law:
$dN/dE \propto E^{-\alpha}$, where $dN/dE$ is the number
of photons per energy interval.
We made simulations of fully polarized GRBs with power
law indices $\alpha = 2.4$, $2.6$, and $2.8\,$.
These simulations were made with uniform time distribution, 
corresponding to a constant flux during the burst.
Between 150 keV and 2 MeV, a power law index of $\alpha = 2.6$ gives 
a detected spectrum that agrees well with the one of GRB021206,
see Fig.\ \ref{fig:comp_spec}.

\begin{figure}
\plotone{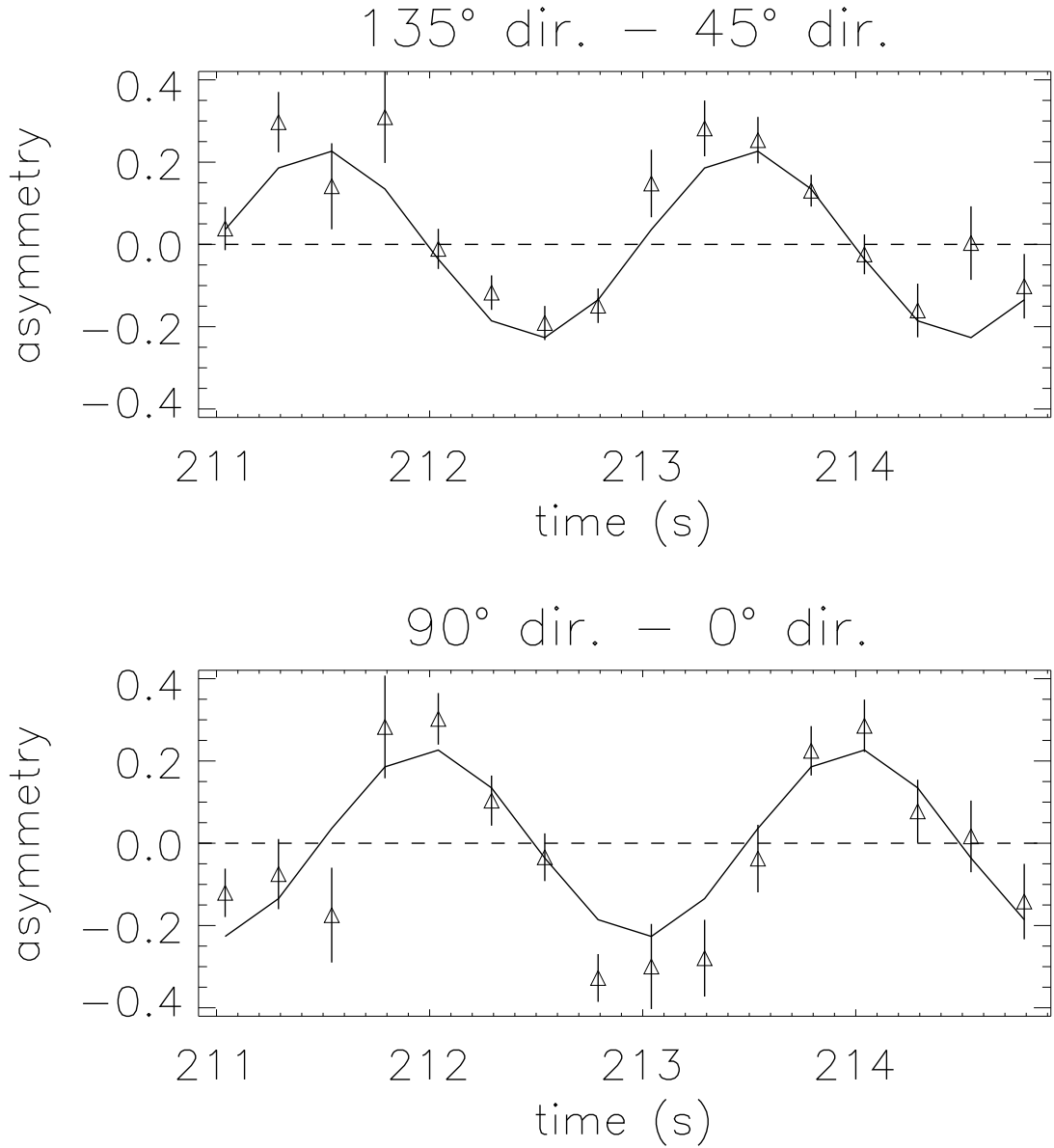}
\caption{Similar to Fig.\ \ref{fig:asym}, but for simulated data.
   The simulation  includes decimation and the light curve 
   variability. The time offset is arbitrary, and $T_{rot,sim}=4.0\,$s.
   The fitted amplitude is given in Eq.\ (\ref{eq:my100}). 
   \label{fig:cdiff_sim}}
\end{figure}

We also made a high-statistics simulation of a fully polarized
GRB with $\alpha = 2.6$ 
taking into account the highly variable light curve of GRB021206.
After applying decimation, 
we found $N_{C,sim} = 4203\, (\pm 82)$ Compton scattering
candidates.
The asymmetry plot for this simulation
is shown in Fig.\ \ref{fig:cdiff_sim}. 
The fitted modulation amplitude is 
\begin{equation}   \label{eq:my100}
 \mu_{100}= (21.0 \pm 2.7)\mbox{\%}.
\end{equation}
The fit result for the phase, $\phi = (87.2 \pm 7.5)\arcdeg$,
is in good agreement with the simulated value of $82.7\arcdeg$.
The $\chi^2$ in the fit is $28.9$ (28 DoF).
The error bars have different length because of the
statistical errors of the underlying light curve.
A uniform time distribution gives error bars of similar length.

\begin{figure}
\plotone{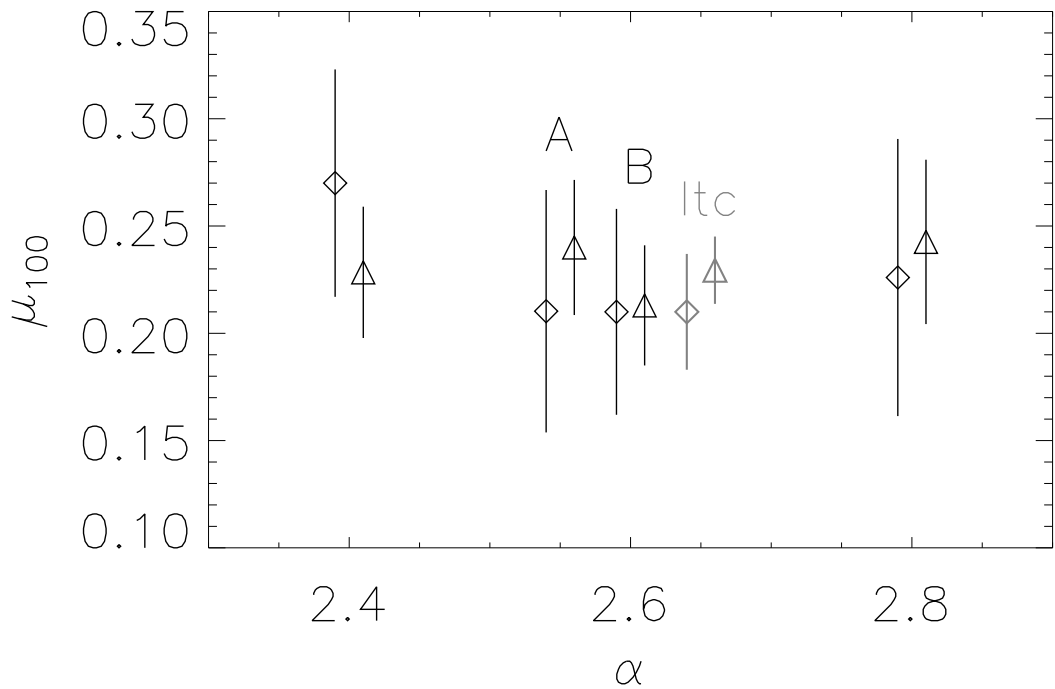}
\caption{Modulation amplitude $\mu_{100}$ of a 100\% polarized GRB for
  different power law indices $\alpha$. 
  Diamonds: the simulated data were decimated.
  Triangles: simulated data without decimation.
  At $\alpha=2.6$, we made two simulations with uniform time
  distribution ('A' and 'B') and one taking into account the light curve
  variability ('ltc', see also Fig.\ \ref{fig:cdiff_sim}
  and Eq.\ (\ref{eq:my100})).     
   \label{fig:simresult}}
\end{figure}

The  fitted asymmetry modulations $\mu_{100}$ are plotted 
as a function of $\alpha$ in Fig.\ \ref{fig:simresult}.
At $\alpha=2.6$ we have two simulations with uniform 
time distribution (called 'A' and 'B')
and one taking into account the light curve variation (called 'ltc').
The values obtained by including decimation are marked
with diamonds, undecimated results are marked by triangles.
The mean value of the 
uniform simulations, with decimation, is
$\mu_{100}= (22.9 \pm 2.7)$\%, in good agreement with 
Eq.\ (\ref{eq:my100}).

We conclude from Fig.\ \ref{fig:simresult}:
(i)~Within statistical errors, 
decimation does not affect the modulation amplitude $\mu_{100}\,$,
(ii)~$\mu_{100}$ does
not depend significantly on the power law index $\alpha$ in the 
range 2.4 -- 2.8,
(iii)~$\mu_{100}$ does not depend on the light curve variation.
This can be understood because each
time bin can be thought of as
an independent experiment. Whether they are simulated with
more or less statistics should not bias the result.  
 
The simulations also show that less than half of the so-called
Compton scattering candidates are due
to the process we want to analyze: an incoming
photon undergoes its first interaction within a
RHESSI detector and is then Compton scattered into 
a second detector, without additional intervening interaction.
Some photons undergo a first interaction before they
reach a germanium detector (of the order 20\%).
Other photons are scattered twice in the first detector
before being detected in a second detector (of the order 40\%).  

We also made simulations with unpolarized photons and 
a power law index $\alpha = 2.6$.
We obtain (decimation is applied)   
$\mu_{unpol} = (5.3 \pm 5.4)\,$\% 
for uniform time distribution
and $\mu_{unpol} = (2.5 \pm 2.7)\,$\% for a simulation 
taking into account the light curve of GRB021206.
Both fits agree with $\mu_{unpol}= 0$ within statistics. 

Simulations of GRB030519B gave comparable values for the
modulation amplitude $\mu_{100}$ as the simulations of GRB021206.

The fraction of accidental coincidences was chosen to be small in 
these simulations. A high fraction of accidental
coincidences would increase the statistical errors.

\subsection{Distant detector pairs}
\label{sec:distant}
As a test, we determined the number of Compton 
scattering candidates as described in section \ref{sec:c_types},
for the 17 distant detector pairs instead of the close
detector pairs. 
From {\em simulations} we obtain a ratio
\begin{equation}  \label{eq:ratio-dc} 
 N_{C,dist}/N_{C,close} \approx (11 \pm 1) \mbox{\%}\;.
\end{equation} 

With the data of {\em GRB021206}, we obtain for the 
distant detector pairs:
$N_{tot} =1007\, (\pm 32)$,
$N_{acc} = 839\, (\pm 9)$,
$N_{BG} =  154\, (\pm 9)$,
and therefore  $N_{C,dist,obs} = 16\, (\pm 34)$.
Combining Eq.\ (\ref{eq:N-C-grb021206}) with Eq.\ (\ref{eq:ratio-dc}),
we would expect $N_{C,dist,ex}=85\, (\pm 9)$. This is
slightly higher than the observed number.

With the data of {\em GRB030519B}, we obtain for the 
distant detector pairs:
$N_{tot} =566\, (\pm 24)$,
$N_{acc} =147\, (\pm 4)$,
$N_{BG} = 345\, (\pm 13)$,
and therefore  $N_{C,dist,obs} = 74\, (\pm 28)$.
Combining Eq.\ (\ref{eq:N-C-grb030519b}) with Eq.\ (\ref{eq:ratio-dc}),
we would expect $N_{C,dist,ex}=68\, (\pm 8)$. This is 
in excellent agreement with the observed number.

The results show that the
distant pairs do not contribute significantly to the
number of Compton scattering candidates.

\section{DISCUSSION}

\subsection{Summary of our method and results}
We select and identify coincidences according to
the parameters given in Sect.\ \ref{sec:coinc} and \ref{sec:c_types}.
The efficiencies of the cuts used are listed in Table \ref{tbl:cut_eff}.
These narrow cuts lead to a comparably small error for the
number of Compton scattering candidates (see section
\ref{sec:comp}).

By comparing scattered events in two directions that
make a relative angle of 90$\arcdeg$, we eliminate most
systematic effects induced
by the strongly varying light curve.
The asymmetry plot we present in Fig.\ \ref{fig:asym}
is entirely based on observational data;
simulations are only necessary to find the possible
degree of polarization.

When processing simulated data of a fully polarized
GRB with our method, we find
a modulation $\mu_{100}$ of the order 20 to 25\%
(see Fig.\ \ref{fig:simresult}).
The observed value $\mu_{p}$ has to be compared
with the simulated value $\mu_{100}$ to obtain
the polarization degree $\Pi_{GRB}$ of a GRB:
\begin{equation}  \label{eq:Pgrb}
  \mu_{p} = \Pi_{GRB} \cdot \mu_{100} \;.
\end{equation}
From theoretical considerations (see Sect.\ \ref{sec:intro}),
a maximum polarization of $\Pi_{max} \approx 0.7$ is
expected. Thus, the maximum observable
modulation amplitude $\mu_{max}$ is of the order 15\%.

The data of both GRBs presented, namely GRB021206 and GRB030529B,
agree with zero polarization (see Eq.\ (\ref{eq:mp-grb021206})
and Eq.\ (\ref{eq:mp-grb030519b})).
But given the large  error bars 
(of the order 9\% for both GRBs), 
the maximum modulation of about 15\% cannot be excluded.

The formula (\ref{eq:Pgrb}) can be written in the form
$\Pi_{GRB} = \mu_{p} / \mu_{100}$.
In the case of GRB021206
we obtain a mean value of $\Pi_{GRB}=41$\%
for the polarization degree
(using Eq.\ (\ref{eq:mp-grb021206}) and Eq.\ (\ref{eq:my100})).
However, proper error propagation is complicated in this case. 
A simple estimate can be obtained by using
the 1$\sigma$ limits of $\mu_{p}$ and  $\mu_{100}$:
$\Pi_{GRB,min}= (8.6-9.6)/(21.0+2.7)=-3$\% and
$\Pi_{GRB,max}= (8.6+9.6)/(21.0-2.7)=98$\%.
Therefore, the result of the polarization analysis can
be summarized as 
\begin{equation}
  \Pi_{\mbox{\scriptsize GRB021206}} = (41^{+57}_{-44}) \mbox{\% .}
\end{equation}

\subsection{Discussion of our method}
\label{sec:our-disc}
 
The way we treat multiple coincidences and 
how we estimate the number of accidental coincidences
is not fully consistent. If the photon flux is so high 
that the chance for triple accidental coincidences is not negligible,
our method underestimates the number of Compton scattering
candidates systematically. 
When we determine
$N_{tot}$ we do not accept any triple coincidences --
including triple {\em accidental} coincidences --,
but afterwards we again subtract the contribution from
triple accidental coincidences, thus 
obtaining a number $N_C$ that is too low. 
This underestimate is smaller, the narrower the cuts  for accepting 
coincidences. That is the reason why we apply the
no-multiples cut as last cut.

In the case of GRB021206, 
most of the triple coincidences are physical.
But some triple coincidences are purely accidental,
and a comparable number of triple coincidences are mixed, i.e.  
a real coincidence is accidentally accompanied by an event
in a third detector. 
We estimate that the real number of Compton scattering candidates
would be of the order 50 
coincidences higher than 
the number given in Eq.\ (\ref{eq:N-C-grb021206}).
An effect of the same order is expected for the
distant detector pairs. The numbers cited in Sect.\ \ref{sec:distant}
for GRB021206 show that the observed number ($N_{C,dist,obs}$) of Compton
scattering candidates for the distant pairs
is $\approx$70 coincidences less than the expected number ($N_{C,dist,ex}$).
This discrepancy is indeed of the order 50.
To determine this number exactly is beyond the scope of this
work, especially since it would only slightly affect the result of the 
polarization analysis and 
does not affect the conclusion. The estimate given above is
fully sufficient for our purposes.

In the case of GRB030519B, 
the chance for triple accidental coincidences is very small. 
The number given in 
Eq.\ (\ref{eq:N-C-grb030519b}) is therefore correct.
This is further confirmed by the numbers cited in Sect.\ \ref{sec:distant},
where $N_{C,dist,obs}$ and $N_{C,dist,ex}$
agree very well for GRB030519B.

Another point to discuss is the 
normalization we use in Eq.\ (\ref{eq:norm_cs_ltc}), and 
the assumption that the efficiencies of the four scattering
directions are proportional to each other.
If statistics were so good, that this approximations could not be
used, the $\chi^2$ of the fit of the asymmetry plot would be too high.
But even with the high-statistics
Monte Carlo simulations we have performed,
our fits of the asymmetry plot gave acceptable results.
Another advantage of our Eqs.\ (\ref{eq:norm_cs_ltc}) and 
(\ref{eq:asym}) is that the so defined asymmetry 
does only depend on data and thus cannot suffer from
systematic errors of the simulations. 
We conclude that our approximations are good enough
within the accuracy necessary for this work.

\subsection{Systematic effects}
 \label{sec:systemtic}
 
We note that the light curve of GRB021206 has
two maxima around 22:49:15 UT and around 22:49:17 UT.
They are separated by roughly 
half a rotation, i.e.\ by the period that a possible
polarization signal would have.
Effects that are related to the count rate
can therefore possibly be mistaken for a polarization signal. 

The accidental coincidences on the one hand,
 are proportional to the 
square of the total GRB light curve: $N_{acc}(t) \propto N_0(t)^2$.
If accidental coincidences are mistaken for 
GRB induced real coincidences and normalized with the
light curve, an effect $\propto N_0(t)$ would be observed. 

Background coincidences on the other hand,
are constant with time (at least approximately).
If background coincidences are mistaken for GRB induced
real coincidences and normalized with the light curve, 
this would result in an effect $\propto 1/N_0(t)$.
We observed for example that
when we did not subtract the background coincidences, the
coincidences from distant detectors did not cancel out 
but showed a significant modulation.

Since we identify the Compton scattering candidates carefully,
our analysis should not suffer from these effects.
By comparing simultaneous scattering rates of perpendicular
directions, we further minimize systematic effects related
to the light curve.

\subsection{Comparison with previous works}
\label{sec:comp}
In Table \ref{tbl:comp} we compare our number of accepted
coincidences with the results presented by 
\citet{CB2003} (hereafter CB) and \citet{RF2003} (hereafter RF).
We would like to emphasize that the total number
of coincidences ($N_{tot}$) depends {\em strongly} on
the cuts used. 
The number of Compton scattering candidates
($N_C$) is less dependent on the cuts used. 
Its error $\sigma_{N_C}$, however, 
depends again on $N_{tot}$, 
namely $\sigma_{N_C} > \sqrt{N_{tot}}$, see Eq.\ (\ref{eq:nC}).

We cannot confirm the high number of Compton scattering candidates
found by CB, our value being more than a factor 10 lower. 
On the other hand, we confirm the result presented by RF,
although our error is 3 times smaller.

\subsubsection{Comparison with CB}
\label{sec:comp_cb}
We can obtain a high
number of coincidences ($N_{tot}$) by widening 
the acceptance cuts, see first line of Table \ref{tbl:cut_eff}.
But by widening the cuts we accept much more 
accidentals, but only a few more
Compton scattering candidates. However,
we cannot reproduce CB's  high number $N_{tot}$ and,
at the same time, their relatively small
$N_{acc}/N_{tot}$-ratio.

The cuts used in CB are not described by the authors.
A few details became very recently available in \citet{CB_slac}.
In Table \ref{tbl:comp} we use the following cuts
in order to reproduce CB's result: 
(i) time interval $14.75\,\mbox{s} \leq t \leq 19.75\,\mbox{s}$, 
(ii) as energy cut: $30\,\mbox{keV} \leq E_k$ and 
   $150\,\mbox{keV} \leq E_i + E_j \leq 2000\,\mbox{keV}$, 
(iii) as $dt$-cut: $|dt| \leq 3\,$b$\mu$s, 
(iv) no-multiples cut; 
the kinematical cut is not applied and 
all detector pairs are accepted. 
Our slightly higher $N_{tot}$ is probably
due to a different treatment of multiple coincidences.

In order to extract the polarization signal, CB rely strongly
on the simulation of a non-polarized GRB.
As a simple test of how well simulations and data agree, 
we can use the ratio of Compton scattering candidates to the
total 0.15-2.0$\,$MeV light curve events:
$ r_{C} =  N_{C}/N_{0.15-2.0}$.
Since $N_C$ depends on the cuts used, $r_C$ does as well.
We use simulations with a low fraction of accidental coincidences
in order to minimize systematic effects.
Such simulations give values around 
$r_{C,sim} \approx 1.33$\%, 
depending slightly on the power law index $\alpha$,
namely $d r_C /d \alpha  \approx -0.35$\%.
Our data analysis gives $r_{C,dat} \approx 1.2$\%,
a value slightly smaller than from simulations. 
The likely reason is described in Sect.\ \ref{sec:our-disc}.
If the chance of triple accidental coincidences is not
negligible, as in the case of GRB021206, $N_C$ is underestimated,
and therefore $r_{C,dat}$ as well.
We conclude that $r_{C,dat}$ and $r_{C,sim}$ agree 
reasonably well using our method and simulations.
If we use similar cuts to CB (see last paragraph),
we obtain $r_{C,sim} \approx 1.9$\%.
From Fig.\ 1  of CB we can estimate the
total number of GRB photons in the 150 to 2000 keV band
to be $N_{0.15-2.0} \approx 71\,000$ (in agreement
with the number obtained from the event list).
CB obtain $N_{C}=9840$ Compton scattering candidates,
resulting in $r_{C,CB} \approx 14$\%, i.e. in 
a value 7 times higher than expected from simulations.
This could be another indication that CB's number of Compton
scattering candidates is too high.

If CB's high number of Compton scattering candidates was
partly due to accidental coincidences, the effect discussed
in Sect.\ \ref{sec:systemtic} were possibly responsible for  
the polarization signal found by CB.

CB's value for the modulation factor, $\mu_{100} = (19 \pm 4)$\%, 
which they obtained from simulations and analytical estimates, 
agrees very well with our value, if we omit the 
kinematical cut (\ref{e_special}). In that case, we obtain
$\mu_{100} = 18.5$\% with a statistical error of $\pm 2.5$\%.

\subsubsection{Comparison with RF}
 
The main differences between our and RF's analysis in
determining the number of Compton scattering candidates are the
following:
\begin{enumerate}
\item Instead of the $dt$-cut (\ref{dt_cut}), 
      RF mention to use 
      $\Delta T \leq 5$ b$\mu$s.
      The quite high rate at $\Delta T=4\,$b$\mu$s in Fig.\ 2 of RF
      is interpreted by them 
      as probably due to real coincidences, whereas it is caused
      by the decimation procedure (see Fig.\ \ref{fig:dtplot}).\\
      Our very narrow $dt$-cut is one of the main reason for our
      much smaller $N_{tot}$ and $N_{acc}$. 
\item Instead of accepting only close detector pairs
      RF accept also distant pairs.
      Accepting distant detector pairs introduces
      mainly background noise (see section \ref{sec:distant}).
\item We determine the number of accidental coincidences from
      a time shifted coincidence window.
      This has the advantage of being independent of the high
      variability of the GRB light curve. Different
      efficiencies of the detectors and effects due to
      the decimation procedure are also taken into account. 
      By making the time window for the time shifted coincidences
      wide, we obtain a relatively small statistical error in $N_{acc}$.
\item Instead of our energy cut (\ref{E0_cut}), RF use 
      150 keV $\leq E_k \leq$ 2000 keV.
      This narrower energy cut  eliminates real coincidences.
\item RF replace front/rear coincidences within the same 
      detector by a single event at the earlier time stamp.
      Photons that made a first interaction in the front segment
      before being Compton scattered in the rear segment into
      another detector mostly lost their polarization information
      in the first interaction. Such events do not contribute
      to the polarization signal.      
\item RF use the same $5\,$s long time interval as CB, whereas
      we use $4.09\,$s (= one full rotation).  
\end{enumerate}
We also tried to reproduce RF (see Table \ref{tbl:comp})
by using the following cuts in our analysis:
(i) time interval $14.75\, \mbox{s} \leq t \leq 19.75\, \mbox{s}$,
(ii) as energy cut: $150\, \mbox{keV} \leq E_i \leq 2000\, \mbox{keV}$,
(iii) as $dt$-cut: $|dt| \leq 3\, $b$\mu$s,
(iv) no-multiples cut;
all detector pairs are  accepted and
no kinematical cut is applied.
Since RF used $|dt| \leq 4\, $b$\mu$s as $dt$-cut (presumably, the 
meaning of RF's $\Delta T$ is not entirely clear) -- 
a cut not easily feasible with our program, that takes into account
the repeating 4\,b$\mu$s-structure caused by decimation --
we added $2/7 \cdot N_{acc}$ to $N_{tot}$ and $N_{acc}$ 
in order to estimate the effect of the wider $dt$-cut.
RF's slightly higher value for $N_{tot}$ than our
reproduction  can be explained by point 5 above.

RF's polarization analysis  is not clear to us.
Their basic Eq.\ (4) is valid for Compton scattering coincidences.
However, it is used for all coincidences, including
about 80\%  accidentals, for which
an expression like
$ N_{i,j}(t)=c \cdot N_i(t) \cdot N_{j}(t)  $
would be more correct.
We further do not understand how to get from RF's equation
(7) to equation (9).

Concerning RF's Section 7, we agree with RF that the 
uncertainty in CB's simulation of an unpolarized GRB is
not discussed in CB. But, in our view, the uncertainty in the 
simulation of an unpolarized GRB and the uncertainty in
the simulations of $\mu_{100}$ are not as closely related as RF claim.

A further evidence that the number of accidental
coincidences is much higher than presented in CB 
comes from the analysis of RF.
RF attempted to reproduce CB although they had to estimate CB's cuts.
In Fig.\ 7b of RF, they 
reproduced the ``angular light curve'' of CB by using a narrower
energy cut than CB, but accepting $\Delta T \leq 8$b$\mu$s.
This clearly increases only the number of  accidental coincidences.
/bin/sh: This: command not found

The fact that RF were able to reproduce CB's angular light curve
by using very different cuts can be understood,
if both sets of cuts, CB's and RF's,
mostly accept accidental coincidences
(R.E.\ Rutledge, private communication).

\subsection{Properties of a GRB that can be searched for
          polarization with RHESSI}

\label{sec:req}
A search for GRB polarization using RHESSI data 
is feasible only for very strong GRBs.
A fully polarized
GRB would yield a modulation amplitude $\mu_{100} \approx 0.21$.
In order to distinguish  
between {\em no} polarization and {\em full} polarization,
the error $\sigma_{\mu_p}$ of the fitted modulation amplitude should be
around 0.05 or smaller.

For both GRBs presented in this article  
we find $\sigma_{\mu_p} \approx 0.09$ (see Eqs. (\ref{eq:mp-grb021206}) 
and (\ref{eq:mp-grb030519b})). 
At first sight, it may seem surprising that both errors are similar,
even though the 25-100 keV fluence of GRB021206 as measured by Ulysses 
was more than 5 times higher than for GRB030519B
(see \citet{grb021206} and \citet{IPN}).
The main reason is the decimation of the RHESSI data. 
If decimation had not been active
during GRB021206, RHESSI would have observed 
about 2.8 times more Compton scattering candidates,
about 2.8 times more accidental coincidences, and
about 2.2 times more background coincidences.
Thus, the error $\sigma_{\mu_p}$ would be about a factor 
$\sqrt{2.8}$ smaller, i.e.\  $\sigma_{\mu_p} \approx 0.06$.
This would be just above the limit to distinguish 
between {\em no} polarization and {\em full} polarization.
Another reason for the similar $\sigma_{\mu_p}$ 
is the better $N_{C}/(N_{acc}+N_{BG})$  ratio
in the case of GRB030519B, namely 0.72 compared
to 0.56 for GRB021206.
The error $\sigma_{N_C}$, and therefore the error $\sigma_{\mu_p}$,
depends on the total number of coincidences: 
$\sigma_{N_C} > \sqrt{N_{tot}}$, see Eq.\ (\ref{eq:nC}).
In the case of GRB021206, $N_{tot}$ is 
dominated by accidental coincidences, while for GRB030519B
$N_{tot}$ is dominated by background coincidences. 
The optimal case, however, would be if
$N_{tot}$ were dominated by Compton scattering candidates.

A GRB suitable for polarization analysis with RHESSI
should have a fluence comparable to GRB021206 and the 
data should not be decimated. Furthermore, it should 
last about $10\,$s 
in order to be neither dominated by accidental coincidences
nor by background coincidences.
For such a GRB,
a polarization degree of 100\% would be distinguishable from
zero polarization at a confidence level of several standard deviations.
Since GRB021206 was one of the strongest GRBs ever observed,
the probability for such a GRB to occur is small,
given that it should also come 
from a direction close to RHESSI's rotation axis.

\section{CONCLUSIONS}

The possibility that prompt GRB emission is strongly linearly polarized
at $\gamma$-ray energies deserves attention with the advent of detectors that
can potentially measure such polarization. Polarization carries
fundamental information on the  orientation of magnetic fields in
the source and eventually helps confine the magnetic field
generation mechanism. Its detection is difficult due to various
interactions in presently available  detector systems. The RHESSI
satellite can in principle be used as a polarimeter by using
the direction dependence of Compton scattering in conjunction with the 
rotation of the satellite. The effects are subtle, however, and
require accurate knowledge of the mass distribution of the satellite
and the detector geometry. 
The methods essentially analyze Compton scattered GRB photons 
that induce coincidences in detector pairs.
A principal difficulty is the separation
of such events from accidental and background coincidences
in the same detector pairs.

Two previous publications \citep{CB2003, RF2003} have addressed this problem,
arriving at essentially opposite conclusions for the same observed GRB. 
Whereas \citet{CB2003}
claim to detect maximum polarization ($\Pi = 80\pm 20$\%), 
\citet{RF2003} challenge this
result and find no significant constraint for the polarization degree.

We have revisited
the problem of GRB polarization measurements with the RHESSI satellite. 
Our basic test case is GRB021206 used by
\citet{CB2003} and \citet{RF2003} for
their respective analyses. 
By applying well justified selection criteria
for coincidences in  energy, time, and scattering angle,
we found $N_{C} = 770 \pm 49$ Compton scattering candidates.
By plotting these 770 events as ``coincidence light curves''
for the four different scattering directions, 
we could define an asymmetry and search for a 
possible polarization signal.
We compare the result of our polarization analysis 
with simulations of a 100\% polarized GRB. 
We cannot reject the null hypothesis that
the burst is unpolarized, but neither can we 
significantly detect any non-zero
polarization. The maximum possible synchrotron 
polarization degree would, for
the measured spectral index of the burst, be of order 70\%,
fully compatible with the data.
We conclude that
{\em no statement on $\gamma$-ray polarization 
can be made for GRB021206}.

Our result contradicts the statements of \citet{CB2003}, who
find more than 9000 Compton scattering candidates ($N_C$) and
claim to see a polarization signal at the 5.7$\sigma$ level.
The main problem in their analysis is the number
of accidental coincidences ($N_{acc}$) that is, in our view, 
not determined correctly.
\citet{CB2003} obtain their signal by comparing a measured 
``angular light curve''
with simulations of an unpolarised GRB. If they do not use a
correct $N_{acc}/N_C$-ratio in their simulation,
then the presented simulated data points cannot be trusted.

Our result agrees in many aspects with the reanalysis presented by
\citet{RF2003}. By making a more sophisticated coincidence
selection, we obtain much smaller errors, however. 
A few points of caution in
the polarization analysis of \citet{RF2003} have been mentioned in
our presentation.

We also analyzed the strong GRB030519B but found 
fewer Compton scattering candidates than for GRB021206 
and, again, no statement about the polarization
degree of GRB030519B can be made.

We therefore conclude that RHESSI has not yet measured the degree of
polarization of any observed GRB. Implications on magnetic field orientation
based on RHESSI results are therefore premature, and the physics of
magnetic field generation and structure in the $\gamma$-ray source of GRB
must rely on alternative information for the time  being. 
On the other hand, our analysis suggests that at least in principle,
RHESSI does  offer the capability of measuring polarization.
A GRB with comparable fluence as GRB021206, but lasting 
$\approx 10\,$s, would be required 
to significantly distinguish maximum polarization of a GRB
from zero polarization. In addition, a GRB suitable
for polarization analysis should come from a direction close to
RHESSI's rotation axis.



\acknowledgments
We thank O.\ Wigger, A.\ Mchedlishvili  
and E. Kirk  for many
helpful discussions and encouragement.
We also thank our referee, D.M.\ Smith 
(SCIPP,  University of California, Santa Cruz), 
for careful reading of our manuscript and  
his constructive comments.




\clearpage


\clearpage

\begin{deluxetable}{crrr}
\tablecaption{Counts per detector segment
         \label{tab:counts}}
\tablewidth{0pt}
\tablehead{
\colhead{detector} & 
\colhead{front}   & 
\colhead{rear}   &
\colhead{total} 
}
\startdata
 1    &  6971 &  8147 &  15118 \\
 2    & 24788 &     0 &  24788 \\
 3    &  6274 &  7387 &  13661 \\
 4    &  5782 &  8280 &  14062 \\
 5    &  6676 &  6523 &  13199 \\
 6    &  4306 &  7482 &  11788 \\
 7    &  5293 &  7500 &  12793 \\
 8    &  6106 &  7850 &  13956 \\
 9    &  3651 &  7797 &  11448 \\ \hline
total & 69847 & 60966 & 130813 \\ 
\enddata
\tablecomments{Number of counts in each detector, time and energy
         interval as indicated in Fig.\ \ref{fig:ltc} and
         Fig.\ \ref{fig:rspec}.}
\end{deluxetable}

\clearpage
\begin{deluxetable}{lcrrrr}
\tablecaption{Number of coincidences, efficiency of cuts 
              \label{tbl:cut_eff}}
\tablewidth{0pt}
\tablehead{
\colhead{applied cuts} & 
\colhead{ref.} &
\colhead{$N_{tot}$}  & 
\colhead{$N_{acc}$} & 
\colhead{$N_{BG}$} & 
\colhead{$N_{C}$}  
}
\startdata
 time interval & Eq.\ (\ref{t0_cut}) & & & &  \\
 $\;\;\;$ and energy cut & Eq.\ (\ref{E0_cut}) & & & &  \\
 $\;\;\;$ and $|dt| \leq 3\,$b$\mu$s  &  &
                    16542 & 13468.8 & 1432.0 & 1641 $\pm$ 141 \\
  $dt$-cut & Eq.\ (\ref{dt_cut})  &  
                     6030 &  3322.6 & 1172.9 & 1535 $\pm\;\,$  83 \\
  close pairs cut  & \S \ref{sec:coinc}  &  
                     4135 &  1868.6 &  790.5 & 1476 $\pm\;\,$  69 \\
  kinematical cut  & Eq.\ (\ref{e_special}) &  
                     2647 &  1093.7 &  512.6 & 1041 $\pm\;\,$  55 \\
  no-multiples     & \S \ref{sec:c_types} &  
                     2141 &  1081.0 &  290.3 &  770 $\pm\;\,$  49 \\
\enddata
\tablecomments{ 
 The total number of coincidences ($N_{tot}$), 
 the number of accidental coincidences ($N_{acc}$), and   
 the number of background coincidences ($N_{BG}$)
 are listed after each additional data cut was applied.
 The corresponding statistical errors are:
 $\sqrt{N_{tot}}$, $\sqrt{N_{acc}/10}$,  
 $\, \approx \! \sqrt{N_{BG}/2}$.  
 The number  of Compton scattering candidates ($N_{C}$) 
 is obtained according to Eq.\ (\ref{eq:nC}).
}
\end{deluxetable}

\clearpage

\begin{deluxetable}{crrlr}
\tablecaption{Number of coincidences, comparison with previous work
              \label{tbl:comp}}
\tablewidth{0pt}
\tablehead{
\colhead{} & 
\colhead{$N_{tot}$} &
\colhead{$N_{acc}$} &
\colhead{$N_{BG}$}  &
\colhead{$N_{C}$}    
}
\startdata   
 CB           & $14916 $
              & $ 4488 \pm 70 $
              & $ 588  \pm\;\, 24 $
	      & $ 9840 \pm\;\, 96$\tablenotemark{a}  \\
 RF           & $\;\, 8230          $
              & $ 6640  \pm  80 $
	      & $  760  \pm 110 $
	      & $  830  \pm 150 $  \\
 present work & $\;\, 2141 $ 
              & $ 1081  \pm  10 $
	      & $  290  \pm\;\,  12 $
	      & $  770  \pm\;\,  49 $ \\  
 ``our CB''
              & $15810   $
              & $13786 \pm    37 $
	      & $  848 \pm\;\, 21 $
	      & $ 1176 \pm   138 $ \\ 
 ``our RF''
              &  $\;\,7788$ 
              &  $6329 \pm 25$
	      &  $ 648 \pm \;\,18 $
	      &  $ 811 \pm \;\,93 $ \\ 
\enddata

\tablenotetext{a}{Error as quoted by CB. From 
    error propagation, this error is expected to be 
    $>\sqrt{N_{tot}}$, i.e.\ $> 122\;$.}
\tablecomments{The total number of coincidences ($N_{tot}$),
    number of accidental coincidences ($N_{acc}$), number of background
    coincidences ($N_{CB}$) and number of Compton scattering candidates
    ($N_{C}$) are compared with previous works. 
    The lines ``our CB/RF''report on our attempts to
    reproduce CB's and RF's numbers by using   
    cuts similar to theirs. 
    See text for details of the cuts used.  }

\end{deluxetable}

\end{document}